\documentclass[prd, reprint, aps, showpacs, longbibliography]{revtex4-1}
 \usepackage[english]{babel}
 \usepackage{amsmath,amsthm,amssymb,amsfonts,mathrsfs, mathtools}
 \usepackage{array}
 \usepackage{graphics}
 \usepackage{aas_macros}
 \usepackage{graphicx}
 \usepackage{hyperref}
 \usepackage{enumerate}
 \usepackage[inline]{enumitem}
 \usepackage{wasysym} 
 \usepackage{bm} 
 \usepackage{cancel}
 \usepackage{float}
 \usepackage[caption = false]{subfig}

 \newcommand{\rmd}{\ensuremath{\mathrm{d}}}
 \newcommand{\rme}{\ensuremath{\mathrm{e}}}

 \newcommand{\f}[2]{\ensuremath{\frac{#1}{#2}} }
 \newcommand{\deriv}[2]{\ensuremath{\frac{\rmd #1 }{\rmd #2}}}
 \newcommand{\sderiv}[2]{\ensuremath{\frac{\rmd^2 #1 }{\rmd #2^2 }}}

 \newcommand{\I}{\ensuremath{\mathbf{\Pi}}}
 \newcommand{\un}[1]{\ensuremath{\mathrm{#1}} }

\begin{document}
 \title{Possible physical realizations of the Tolman~VII solution}
 \author{Ambrish M. Raghoonundun}\email{amraghoo@ucalgary.ca} 
 \affiliation{University of Calgary, 2500 University Drive NW, Calgary, Alberta, Canada, T2N 1N4}
 \author{David W. Hobill} \email{hobill@ucalgary.ca}
 \affiliation{University of Calgary, 2500 University Drive NW, Calgary, Alberta, Canada, T2N 1N4}

 \begin{abstract}
   The Tolman~VII solution for a static perfect fluid sphere 
   to the Einstein equations is reexamined, and a closed form
   class of equations of state (EOSs) is deduced for the first time.
   These EOSs allow further analysis to be carried out, leading to a
   viable model for compact stars with arbitrary boundary mass density
   to be obtained.  Explicit application
   of causality conditions places further constraints on the model, and
   recent observations of masses and radii of neutron stars prove to be
   within the predictions of the model.  The adiabatic index predicted
   is $\gamma \geq 2,$ but self-bound crust solutions are not excluded
   if we allow for higher polytropic indices in the crustal regions of
   the star.  The solution is also shown to obey known stability
   criteria often used in modeling such stars.  It is argued that
   this solution provides realistic limits on models of compact stars,
   maybe even independently of the type of EOS, since most of the EOSs
   usually considered do show a quadratic density falloff to first
   order, and this solution is the unique exact solution that has this
   property.
 \end{abstract}

 \pacs{{04.20.Jb}, {04.40.Dg}, {95.30.Sf}, {97.60.Jd}}
 \maketitle
 \section{\label{sec:intro}Introduction}

 The construction of exact analytic solutions to the Einstein equations
 has had a long history, nearly one hundred years to be more precise.
 However, in spite of the fact that the total number of solutions is
 large~\cite{KraSteMac80} and growing, only a small subset of those
 solutions can be thought of as having any physical relevance.  Most
 solutions exhibit mathematical pathologies or violate simple
 principles of physics (energy conditions, causality, etc.) and are
 therefore not viable descriptions of any observable or potentially
 observable phenomena.

 Indeed, works that review exact solutions and their properties
 demonstrate the difficulties associated with constructing solutions
 that might be relevant to gravitating systems that actually exist in
 our Universe.  Even in the simplest case of exact analytic solutions
 for static, spherically symmetric fluid spheres, it has been shown
 that less than ten percent of the many known solutions can be
 considered as describing a realistic, observable object.  For example,
 \citeauthor*{DelLak98}, using computer algebra methods, reviewed over
 130 solutions and found that only nine could be classified as
 physically relevant~\cite{DelLak98}.  A similar study by
 \citeauthor*{FinSke98} arrived at the same conclusion~\cite{FinSke98}.
 The latter review also introduced an additional criterion that further
 reduced the number of physically relevant solutions to those that have
 exact analytic equations of state (EOSs) of the form \(p = p(\rho),\)
 where \(p\)
 is the fluid pressure and \(\rho\)
 is the matter density.  This class of solutions was called ``the set
 of interesting solutions.''

 In 1939 \citeauthor*{Tol39} introduced a technique for constructing
 solutions to the static, spherically symmetric Einstein equations with
 material fluid sources~\cite{Tol39}.  That method led to eight exact
 analytic expressions for the metric functions, the matter density and
 in some cases the fluid pressure.  Beginning with an exact analytic
 solution for one of the two metric functions, an expression for the
 mass density could be obtained by integration.  With expressions for
 the density and the first metric function in hand, an analytic
 expression for the second metric function could be obtained. This
 often required an appropriate change of the radial variable to obtain
 a simple integral.  All functions could then be written as explicit
 functions of the radial coordinate \(r.\)
 While the fluid pressure could, in principle, be obtained from the
 metric and density functions, Tolman chose not to evaluate the fluid
 pressure in some cases due to the fact that to do so would lead to
 mathematically rather complicated expressions that might be difficult
 to interpret.

 Of the eight solutions presented in his paper, three were already
 known (the Einstein universe, the Schwarzschild--de Sitter solution, and
 the Schwarzschild constant density solution); most of the others
 ``describe situations which are frankly unphysical, and these do have
 a tendency to distract attention from the more useful
 ones.''~\cite{Kin75}.  One, the so-called Tolman~VII solution appeared
 to have some physical relevance, but this was one of the solutions for
 which no explicit expression for the pressure was given.

 The Tolman~VII solution has been rediscovered a number of times and
 has appeared under different names, the
 Durgapal~\cite{DurGeh71,DurRaw79} and the \citeauthor*{Meh66}
 solutions being two examples. That these solutions can be used to
 describe realistic physical systems has been noted by many authors,
 including those of the two review papers mentioned
 above~\cite{DelLak98, FinSke98}.  It has been used as an exact
 analytic model for spherically symmetric stellar systems, and
 additional research has investigated its stability
 properties~\cite{NeaLak01, NeaIshLak01}.  While these later works were
 able to obtain the complicated expressions for the fluid pressure as a
 function of the radial coordinate, according to Finch and
 Skea~\cite{FinSke98} it still was not one of the ``interesting
 solutions'' since it lacked an explicit expression for the equation of
 state.  The choice of parameters that has been taken by different
 authors in order to completely specify the solution in many ways
 prevented the immediate interpretation of the physical conditions
 described by the solution.

 The reasons mentioned above are not sufficient to use or classify the
 Tolman~VII solution as a physically viable one.  Instead, we seek
 physical motivations for the viability of this solution, and indeed we
 find these in many forms:
 \begin{enumerate}[leftmargin=\parindent,labelindent=\parindent,label=(\roman*)]
 \item \label{lameEmden} From a Newtonian point of view, simple
   thermodynamic arguments yield polytropes of the form
   \(p(\rho) = k \rho^{\gamma}\) (here $\gamma$ is the adiabatic index
   sometimes written in terms of the polytropic index $n$, $\gamma = 1 + 1/n$,
 and $k$ is known at the adiabatic constant that can vary from star to star) 
   as viable models for neutron matter.  When coupled with Newtonian
   hydrodynamic stability and gravitation, the result is the Lane-Emden
   differential equation for the density profile, \(\rho(r).\)
   Solutions of the latter, obtained numerically, or in particular
   cases \((\gamma = \infty, 2,\)
   or 1.2) exactly, all have a distinctive density falloff from the
   center to the edge of the Newtonian star.  This is a feature we wish
   physical solutions to have.  Furthermore, this distinctive falloff is
   quadratic in the rescaled radius~\cite{Hor04}, suggesting that even
   in the relativistic case, such a falloff would be a good first
   approximation to model realistic stars, which have a proper
   thermodynamic grounding.

 \item \label{schInt} Looking at viable exact relativistic solutions to
   the Einstein equations, the one used extensively before 1939 and
   even much later, was the Schwarzschild interior solution.  This
   solution has the feature that the density is constant throughout the
   sphere, and is not physical: the speed of sound (pressure) waves in
   its interior is infinite.  However, this solution provides clear
   predictions about the maximum possible mass of relativistic stars in
   the form of the Buchdahl limit~\cite{Buc59}: \(M \leq 4R/9.\)
   The next best guess in this line of reasoning of finding limiting
   values from exact solutions would be to find an exact solution with
   a density profile that decreases with increasing radius, since a
   stability heuristic for stars demands that
   \(\rmd \rho / \rmd r \leq 0,\)
   as expected from~\ref{lameEmden} in the Newtonian case.  Extension
   to the relativistic Lane-Emden equation also requires~\cite{Hor04}
   that \((\rmd \rho / \rmd r)|_{r=0} = 0,\) a property Tolman~VII has.

 \item \label{falloff} Additionally, an extensive
   review~\cite{LatPra01} of most EOSs used from nuclear physics to
   model neutron stars concluded that a quadratic falloff in the
   density is a very close approximation to \emph{most} such nuclear
   models--the differences of drastically different nuclear models
   from Tolman~VII being only minor if only the density profiles were
   compared.  Since Tolman~VII is precisely the unique exact solution
   to the full Einstein field equations that exhibits a quadratic
   falloff in the density profile, we believe that it captures much of
   what nuclear models have to say about the overall structure of
   relativistic stars.
 \end{enumerate}

 These three reasons taken together make a strong case for considering
 the Tolman~VII solution as the best possible exact solution that is
 capable of describing a wide class of EOSs for neutron stars.
 At the very least, it is as good a candidate that captures first-order
 effects in density of \emph{most} nuclear model EOSs, and at best it is
 the model that all realistic nuclear models tend to, while including
 features like self-boundedness naturally, as we shall show.

 The purpose of this paper is to reexamine the Tolman~VII solution by
 introducing a set of constant parameters that we believe provide a
 more intuitive understanding of the physical content of the solution.
 In addition, the solution now becomes a member of the set of
 ``interesting solutions'' since we provide an explicit expression for
 a class of equations of state derived from the solution without any
 further assumptions about the matter, except for the Newtonian-like,
 and physically motivated quadratic falloff of the density.  The EOSs
 will allow for further exploration of the predictions of the solution
 as well as a description of the material that makes up the star. The
 imposition of both the causality conditions where the speed of sound
 in the fluid never exceeds the speed of light and the different
 boundary conditions will provide further restrictions on the
 parameters associated with the solution.  What this all leads to is a
 complete analytic model for compact stars that can be used to compare
 with recent observations of neutron star masses and radii.  That the
 Tolman~VII solution is consistent with all observations of
 astrophysical neutron stars leads to the conclusion that this exact
 solution is physically relevant while having features present in
 compact objects found in nature.

 This article is divided as follows: following a brief historical
 introduction in Sec.~\ref{sec:intro}, we re-derive the Tolman~VII
 solution in Sec.~\ref{sec:pressure}, paying particular attention to
 the pressure expression with physically more intuitive variables.  We
 then invert the density equation and use the pressure expression just
 found to derive a class of EOSs in Sec.~\ref{sec:eos}, where we also
 carry out an analysis of the said class of EOSs.  In the same section,
 we contrast the two different types of physical models that the
 solution admits, and we shall also show how qualitative differences
 arise in the stars' structure and quantitative ones appear in the
 predicted values of the adiabatic indices of the fluid.  Finally we
 provide some brief concluding remarks in Sec.~\ref{sec:conclusion}.

 \section{\label{sec:pressure}The Tolman~VII solution}
 Beginning with a line element in terms of standard areal
 (Schwarzschild) coordinates for a static and spherically symmetric
 metric,
 \begin{equation}\label{eq:metric}
 \rmd s^{2} = \rme ^{\nu(r)} \rmd t^{2} - \rme ^{\lambda(r)} \rmd
 r^{2} - r^{2} \rmd \theta^{2} - r^{2} \sin^{2} \theta \rmd
 \varphi^{2},  
 \end{equation}
 the Einstein equations for a perfect fluid source lead to three
 ordinary differential equations for the two metric variables \(\nu\),
 \(\lambda\),
 and the two matter variables \(\rho\)
 and \(p\).
 However, these variables will not be the most practical ones to carry
 out our analysis.  Instead two related metric functions,
 \(Z(r) = \rme^{-\lambda(r)}\)
 and \(Y(r) = \rme^{\nu(r)/2},\)
 are introduced, as prescribed in Ivanov~\cite{Iva02}.  The reason for
 introducing these new metric variables is that with the assumption
 made for the density function, these variables will transform the
 original nonlinear differential equations into linear ones which may
 then be easily solved.
 \begin{widetext}
 Given the metric equation~\eqref{eq:metric}, the
 Einstein equations reduce to the following set of three coupled
 ordinary differential equations (ODEs) for the four variables
 \(Z,Y,p,\) and \(\rho\):
 \begin{subequations}
   \label{eq:EinR}
   \begin{alignat}{3}
     \label{eq:EinR1}
     \kappa \rho &= \rme^{-\lambda}\left( \f{\lambda'}{r} -\f{1}{r^2}\right) +\f{1}{r^{2}} 
     &&= \f{1}{r^{2}} - \f{Z}{r^{2}} - \f{1}{r}\deriv{Z}{r}, \\
     \label{eq:EinR2}
     \kappa p &= \rme^{-\lambda} \left( \f{\nu'}{r} + \f{1}{r^2}\right) -\f{1}{r^2} 
     &&= \f{2Z}{rY}\deriv{Y}{r} + \f{Z}{r^{2}} - \f{1}{r^{2}},\\
     \label{eq:EinR3}
     \kappa p &= \rme^{-\lambda} \left( \f{\nu''}{2} - 
       \f{\nu'\lambda'}{4} + \f{(\nu')^2}{4} + \f{\nu'- \lambda'}{2r}\right)  
     &&= \f{Z}{Y}\sderiv{Y}{r} + \f{1}{2Y}\deriv{Y}{r}\deriv{Z}{r} + \f{Z}{rY}\deriv{Y}{r} + \f{1}{2r}\deriv{Z}{r}.
   \end{alignat}
 \end{subequations}
 \end{widetext}
 where the primes \((')\) denote differentiation with respect to \(r,\)
 and \(\kappa\) is equal to \(8\pi,\) since in what follows natural units where
 \(G = c = 1\) are introduced.

 The first two equations~\eqref{eq:EinR1}
 and~\eqref{eq:EinR2} can be added together to generate the simpler
 equation\begin{equation}
   \label{eq:EinR1+2}
   \kappa (p+\rho) = \f{2Z}{rY}\deriv{Y}{r} - \f{1}{r}\deriv{Z}{r},
 \end{equation}
 which will be useful later on.  In order to solve this set of ODEs,
 one begins with equation~\eqref{eq:EinR1} and assumes a specific
 functional form for the density, one that is motivated from physical
 considerations according to~\ref{falloff}.  Since this is a linear
 inhomogeneous ODE for $Z(r),$ one can for the appropriately chosen form
 of $\rho(r)$ easily integrate this equation.  The Tolman VII density
 has a simple functional form:
 \begin{equation}
   \label{eq:Density}
   \rho = \rho_{c} \left[ 1 - \mu \left( \f{r}{r_{b}}\right)^{2} \right],
 \end{equation}
 where the constant \(r_{b}\)
 represents the boundary radius as mentioned previously, \(\rho_{c}\)
 represents the central density at \(r=0,\)
 and \(\mu\)
 is a ``self-boundedness'' dimensionless parameter that spans values
 between 0 and 1, so that when it is equal to 0, we have a sphere of
 constant density, and when it is equal to 1, we have a ``natural''
 star, with density vanishing at the boundary.

 Although very simple, this quadratic function is known to provide a
 good approximation for the density profile of a number of neutron
 star's EOSs.  For example, Fig~5 in Ref.~\cite{LatPra01} plots the
 density profile of 12 EOSs and compares them to a function of the
 form given in equation~\eqref{eq:Density} (for the $\mu = 1$ case). 
 Therefore, the claim is that
 this functional form is a generic feature of many different types of 
 nuclear EOSs and this suggests that at the very least some global
 features of such a density profile might describe the bulk properties
 of many compact objects.

 The set of three parameters that describe the density function will
 occur frequently in what follows and will be denoted as:
 \( \I \coloneqq \{\rho_{c}, r_{b},\mu\}.\)
 The form of the density function for \(\mu > 0\)
 is physically realistic, since it is monotonically decreasing from the
 center to the edge of the sphere, as argued previously in~\ref{schInt}
 and~\ref{falloff}, in contrast to the constant-density exact solution
 frequently used to model such objects.

 Additionally, boundary conditions are required for the system, since we
 eventually want to match this interior solution to an external metric.
 Since the vacuum region is spherically symmetric and static, the only
 candidate by Birkhoff's theorem is the Schwarzschild exterior
 solution.  The Israel-Darmois junction conditions for this system can
 then be shown to be equivalent to the following two
 conditions~\cite{Syn60}:
 \begin{subequations}
   \label{eq:Boundary1+2}
   \begin{align}
     \label{eq:BoundaryP}
     p(r_{b}) &= 0, \quad \text{and} \\
     \label{eq:BoundaryZ}
     Z(r_{b}) &= 1-\f{2M}{r_{b}} = Y^{2}(r_{b}),
   \end{align}
 \end{subequations}
 where \(M = m(r_{b})\) is the total mass of the sphere as seen by an outside
 observer, and \(m(r)\) is the mass function defined by
 \begin{equation}
   \label{eq:MassFunction}
   m(r) = 4\pi \int_{0}^{r} \rho(\bar{r}) \bar{r}^{2 }\rmd \bar{r}.
 \end{equation}
 Furthermore, the requirement of regularity for the mass function,
 that it must vanish at the origin of the radial coordinate from
 physical considerations, leads to \(m(r=0) = 0.\) On
 imposing~\eqref{eq:BoundaryZ}, one immediately writes \(Z\) in terms
 of the parameters appearing in the density assumption:
 \begin{equation}
   \label{eq:ZWithParams}
   Z(r) = 1 - \left( \f{\kappa \rho_{c}}{3}\right) r^{2} + 
   \left(\f{\kappa \mu \rho_{c}}{5r_{b}^{2}}\right) r^{4} \eqqcolon 1- br^2+ ar^4.
 \end{equation}
 In contrast, Tolman's method was to assume the second form for $Z$ (or
 equivalently for \(\rme^{- \lambda}\))
 in~\eqref{eq:ZWithParams}, and then obtain the density function
 from~\eqref{eq:EinR1} directly by differentiation.  The physical
 constants \(\mu, \rho_c,\)
 and \(r_b\)
 occur frequently enough in the combinations shown above that the
 constants \(a\)
 and \(b\)
 as defined in~\eqref{eq:ZWithParams} will be used when convenient.
 The solution methods for solving the ODEs obtained from the Einstein
 equations, particularly those leading to the Tolman~VII solution, have
 been given in multiple references~\cite{Meh66,Tol39} and will not be
 reproduced here.

 The complete Tolman~VII solution is specified with the two
 functions~\eqref{eq:Y} and~\eqref{eq:xiCoth} below, together with the
 previously given density function~\eqref{eq:Density}, and the metric
 function \(Z\) in equation~\eqref{eq:ZWithParams}:
 \begin{equation}
   \label{eq:Y}
     Y(\xi) = c_{1} \cos (\phi \xi) + c_{2} \sin (\phi \xi),
 \end{equation}
 where \(\phi = \sqrt{a/4}\).
 The quantity \(\xi\)
 is a new radial variable whose explicit expression in terms of \(r\)
 is
 \begin{equation}
   \label{eq:xiCoth}
   \xi(r) = \f{2}{\sqrt{a}} \coth^{-1} \left( \f{1+\sqrt{1-br^2+ar^4}}{r^2\sqrt{a}}\right)
 \end{equation}
 and it has been employed to simplify the expression of \(Y.\)

 Now that the full solution for the metric functions is known, the
 pressure can be computed through the
 relation~\eqref{eq:PressureMetric}, obtained from a simple
 rearrangement and variable change of~\eqref{eq:EinR1+2}:
 \begin{equation}
     \label{eq:PressureMetric}
     \kappa p(r) =  4 \f{\sqrt{Z}}{Y}\deriv{Y}{\xi} - \f{1}{r}\deriv{Z}{r} - \kappa \rho.
 \end{equation}
 This substitution results in a very complicated-looking expression for the pressure,
 \begin{widetext} 
 \begin{equation}
   \label{eq:PressureR}
   \kappa p(r) = \f{4\phi [c_{2} \cos{(\phi\xi)} - c_{1} \sin{(\phi\xi)}] \sqrt{1 - br^{2} + ar^{4}}}{c_{1}\cos{(\phi\xi)} + c_{2} \sin{(\phi\xi)}} - 4ar^{2} + 2b - \kappa \rho_{c} \left[ 1 - \mu \left( \f{r}{r_{b}}\right)^{2} \right].
 \end{equation}
 \end{widetext}
 So far the two integration constants \(c_{1}\)
 and \(c_{2}\)
 appearing in the expression for \(Y,\)
 and therefore \(p,\)
 are completely arbitrary.  Application of the the boundary conditions
 using equations~\eqref{eq:PressureMetric},~\eqref{eq:BoundaryP},
 and~\eqref{eq:BoundaryZ} leads
 to
 \begin{equation}
 \label{eq:cancel}
 \left. \kappa (p+\rho)
 \right|_{x=x_{b}} = \f{4 \cancel{\sqrt{Z(x_{b})}}}{\cancel{Y(x_{b})}}
 \left. \deriv{Y}{\xi} \right|_{\xi=\xi_{b}} - 2 \left. \deriv{Z}{x}
 \right|_{x=x_{b}},  
 \end{equation}
 where \(x \coloneqq r^{2}\)
 is another radial coordinate, and all the \(b\)-subscripted
 variables are the values at the boundary \(r=r_{b}.\)  The
 cancellation shown results from matching to the exterior Schwarzschild
 solution. However, according to the second boundary
 condition~\eqref{eq:BoundaryP}, the pressure has to vanish at the
 boundary; therefore equation~\eqref{eq:cancel} simplifies
 to\[ \left. \kappa \rho \right|_{x=x_{b}} = \left. 4 \deriv{Y}{\xi}
 \right|_{\xi=\xi_{b}} - 2 \left. \deriv{Z}{x} \right|_{x=x_{b}},\]
 which can be further simplified and rearranged as
 \begin{equation}
   \label{eq:DefAlpha}
   \left. \deriv{Y}{\xi} \right|_{\xi=\xi_{b}} = \f{b-ax_{b}}{4} = \f{\kappa \rho_{c}}{4}\left( \f{1}{3} - \f{\mu}{5}\right) \eqqcolon \alpha.
 \end{equation}
 Since the ODE, equation~\eqref{eq:EinR3} for \(Y,\)
 is second order, a second condition is required.  This is
 simply going to be condition~\eqref{eq:BoundaryZ} restated as
 \begin{multline}
   \label{eq:DefGamma}
   Y(x=x_b) = \sqrt{1-bx_{b}+ax_{b}^{2}} \\= \sqrt{1 - \kappa\rho_{c}r^{2}_{b} \left( \f{1}{3} - \f{\mu}{5}\right)} \eqqcolon \gamma
 \end{multline}
 The two equations~\eqref{eq:DefAlpha} and~\eqref{eq:DefGamma} constitute the
 complete Cauchy's boundary condition on $Y$.  The 
 integration constants \(c_{1}\) and \(c_{2}\) can now be determined
 from the simultaneous equations:
 \begin{align}
 \nonumber
  \left. \deriv{Y}{\xi} \right|_{\xi=\xi_{b}} &= \phi \left[c_{2} \cos{(\phi \xi_{b})} - c_{1} \sin{(\phi\xi_{b})} \right] = \alpha, \\
 \label{eq:BDYsimul} 
 &\therefore c_{2} \cos{(\phi \xi_{b})} - c_{1} \sin{(\phi\xi_{b})} = \alpha / \phi, \\
 \nonumber
  Y(\xi=\xi_b) &= \gamma \\ 
 \label{eq:BDZsimul}
 &\therefore  c_{2} \sin{(\phi\xi_{b})} + c_{1} \cos{(\phi \xi_{b})} = \gamma.
 \end{align}
 This system can be solved by first multiplying~\eqref{eq:BDYsimul} by
 \(\cos{(\phi \xi_{b})}\), and~\eqref{eq:BDZsimul} by \(\sin{(\phi
   \xi_{b})},\) and adding the equations obtained, yielding \(c_{2}.\)
 Similarly, switching the multiplicands and performing a subtraction
 instead yields \(c_{1},\) and these can be given in the form,
 \begin{align}
   \label{eq:C1solved}
   c_{1} &= \gamma \cos{(\phi\xi_{b})} - \f{\alpha}{\phi} \sin{(\phi \xi_{b})}, \\
   \label{eq:C2solved}
   c_{2} &= \gamma \sin{(\phi\xi_{b})} + \f{\alpha}{\phi} \cos{(\phi \xi_{b})}.
 \end{align}
 The
 integration constants are ultimately computed in terms of the parameter set
 \(\I,\)
 and in doing so this completes the specification of the full
 Tolman~VII solution in the new constant scheme.

 An important quantity to consider (since it establishes whether or not
 the solution is relativistically causal) is the adiabatic speed of
 sound waves in the fluid.  The usual definition of this quantity in
 perfect fluids is \(v^{2} = \rmd p / \rmd \rho.\)
 However, it will be convenient to find an expression for the sound
 speed directly from the differential equations, since the expression
 and functional form while completely equivalent is simpler to work
 with.  First, from the expression for the density~\eqref{eq:Density},
 one obtains the
 derivative\[\deriv{\rho}{r} = -\f{2 \mu \rho_{c}}{r_{b}^{2}} r,\]
 which is zero only at \(r = 0.\)
 For the other equation, the conservation of the energy-momentum tensor
 \(\nabla_{i} T^{i}{}_{j} = 0\) reduces to
 \begin{equation}
   \label{eq:TOV}
   \deriv{p}{r} = -\f{\nu'(p+\rho)}{2} = -\f{(p+\rho)}{Y}\deriv{Y}{r},
 \end{equation}
 in the \(j=0\) case.  These two expressions can be used to find \(\rmd p
 /\rmd \rho\) for every value of \(r\) but the center, so that
 \begin{equation}
   \label{eq:SpeedSound}
   v^{2} = \deriv{p}{\rho} = \left( \deriv{p}{r} \middle/ \deriv{\rho}{r} \right) = \f{r_{b}^{2}(p+\rho)}{2\mu\rho_{c}rY}\deriv{Y}{r}.
 \end{equation}
 Using the expressions for all the terms in this formula, one obtains
 a relatively simple expression for the speed of sound. 

 The bulk modulus \(K\)
 of a fluid is a measure of the resistance of a fluid to change its
 volume under an applied pressure.  For perfect fluids it is related to
 the speed of sound in the media through \(K = \rho v^{2}.\)
 This is also a quantity which may be computed for the fluid in the
 interior.  This calculation demonstrates that the order of magnitude of
 the bulk modulus is significantly higher than any currently known substance
 by many orders of magnitude.

\begin{figure}[h!]
 \includegraphics[width=\linewidth]{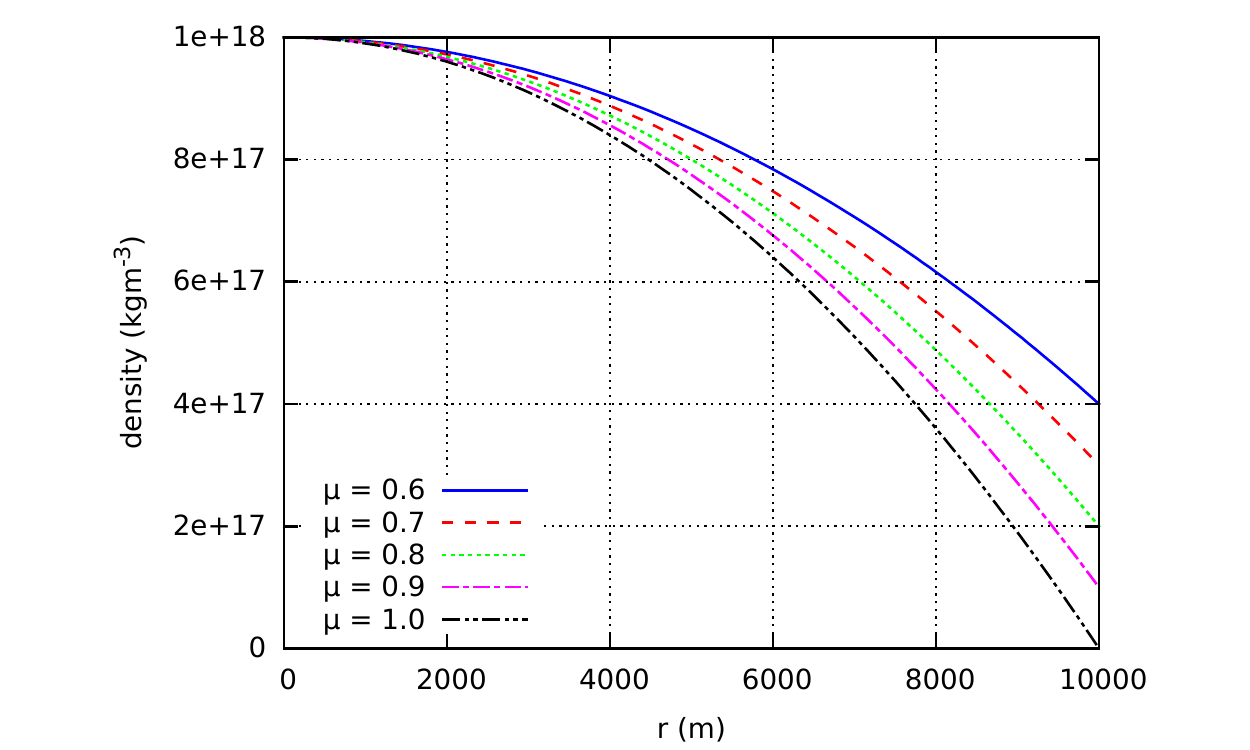}
 \caption{Variation of density with the radial coordinate
   inside the star. The parameter values are $\rho_{c}=\un{1\times
     10^{15} g \cdot cm^{-3}},$ $r_b = \un{1 \times 10^{6} cm}$ and $0.6 \leq \mu \leq 1.0$ .}
 \label{fig:density}
 \end{figure}

 The next step to understanding this solution is to investigate the
 behavior of the solution as the parameter set is varied. The
 particular choice of parameters will be those that are associated with
 what one might expect for realistic compact astrophysical objects.  As
 a result, central densities
 \(\rho_{c} \sim \un{10^{15} g\cdot cm^{-3}}\)
 will be typical. Similarly, radii \(r_{b} \sim \un{10^{6} cm}\)
 (i.e. 10 km) will often be used for the same reason.  As stated above,
 the density profile~\eqref{eq:Density} will decrease quadratically, and
 this provides a good approximation of what one would expect from a
 number of neutron star EOSs.  Figure~\ref{fig:density} plots the density
 as a function of radius for different values of $\mu,$ which controls
 the relation of the surface density to the central density.

 The surface density ranges from a zero
 value when \(\mu=1\)
 to increasingly higher densities as \(\mu\)
 is decreased.  In the literature~\cite{LatPra01}, models having zero
 surface densities have been called ``natural,'' and those with
 nonvanishing surface densities have been called ``self-bound.''  It
 is for this reason that \(\mu\) is called
 the ``self-boundedness'' parameter. 

 \begin{figure}[h!]
 \includegraphics[width=\linewidth]{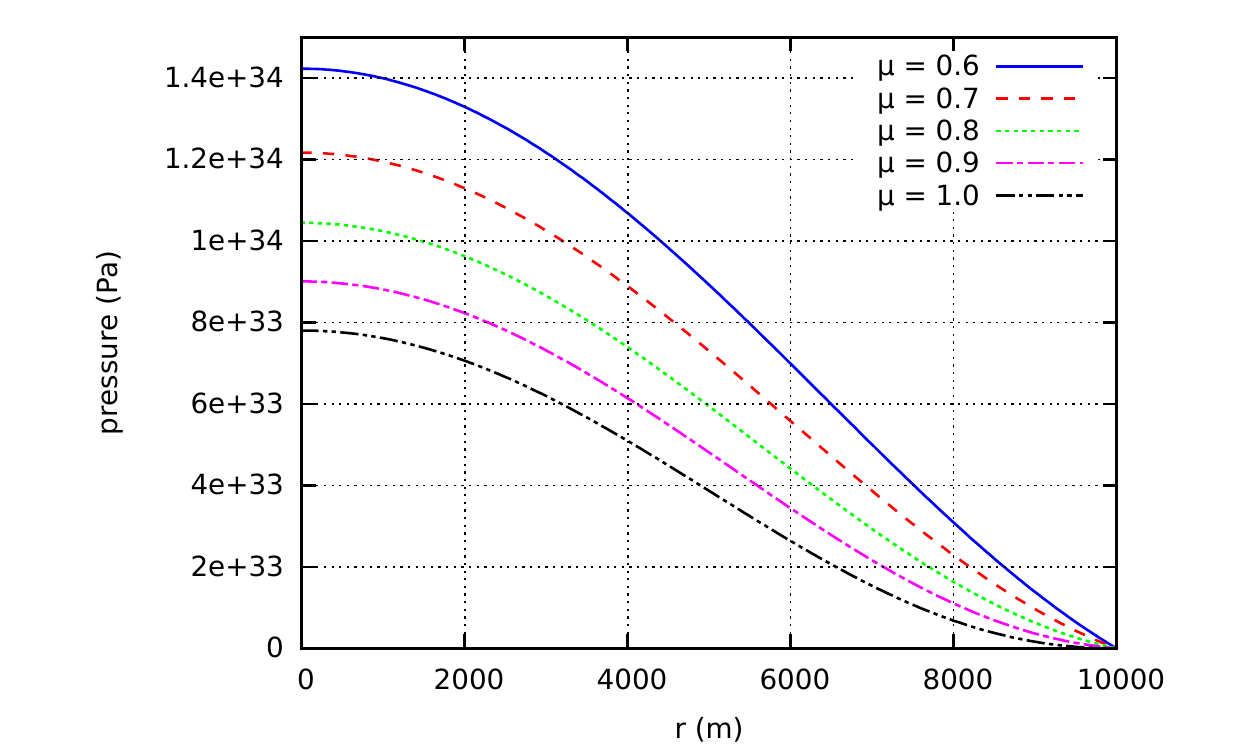}
 \caption{Variation of pressure with the radial coordinate
   inside the star. The parameter values are $\rho_{c}=\un{1\times
     10^{15} g \cdot cm^{-3}},$ $r_b = \un{1 \times 10^{6} cm}$ and $0.6 \leq \mu \leq 1.0$ .}
\label{fig:pressure}
\end{figure}

\begin{figure}[h!]
\includegraphics[width=\linewidth]{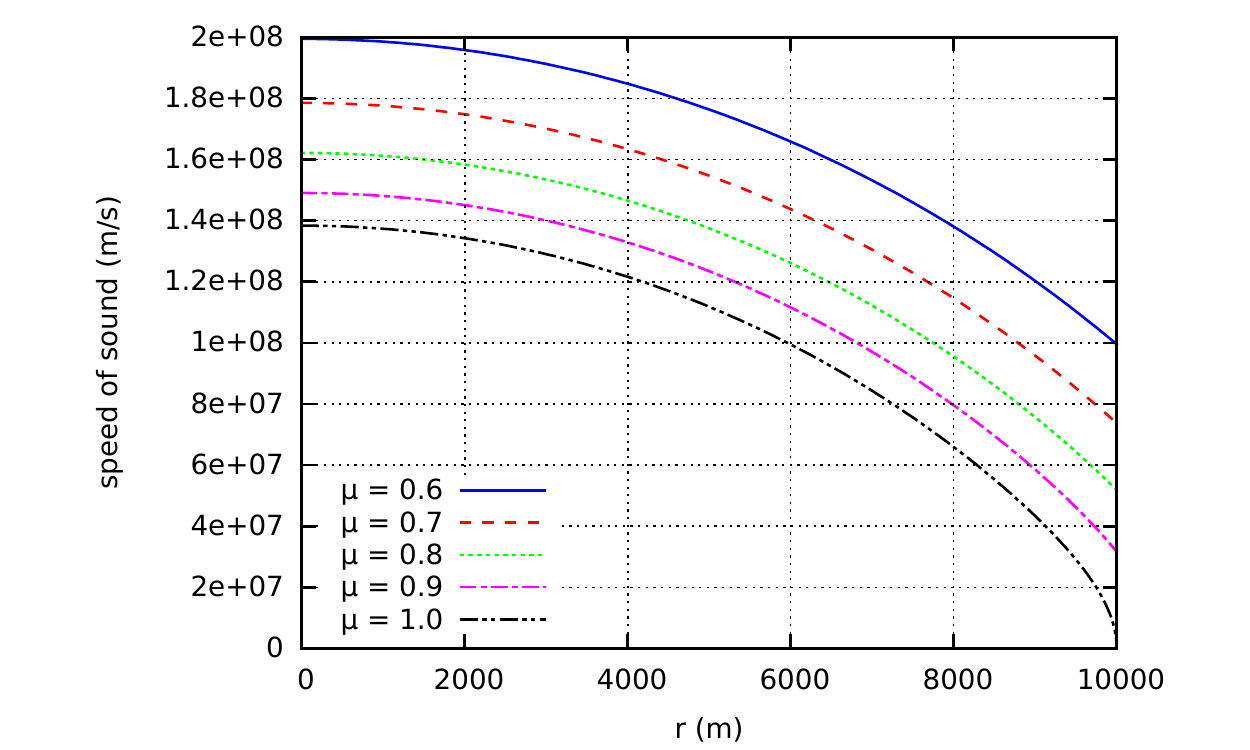}
\caption{Variation of speed of sound with the radial coordinate
  inside the star. The parameter values are $\rho_{c}=\un{1\times
    10^{15} g \cdot cm^{-3}},$ $r_b = \un{1 \times 10^{6} cm}$ and $0.6 \leq \mu \leq 1.0$ .}
\label{fig:sound}
\end{figure}

\begin{figure}[h!]
\includegraphics[width=\linewidth]{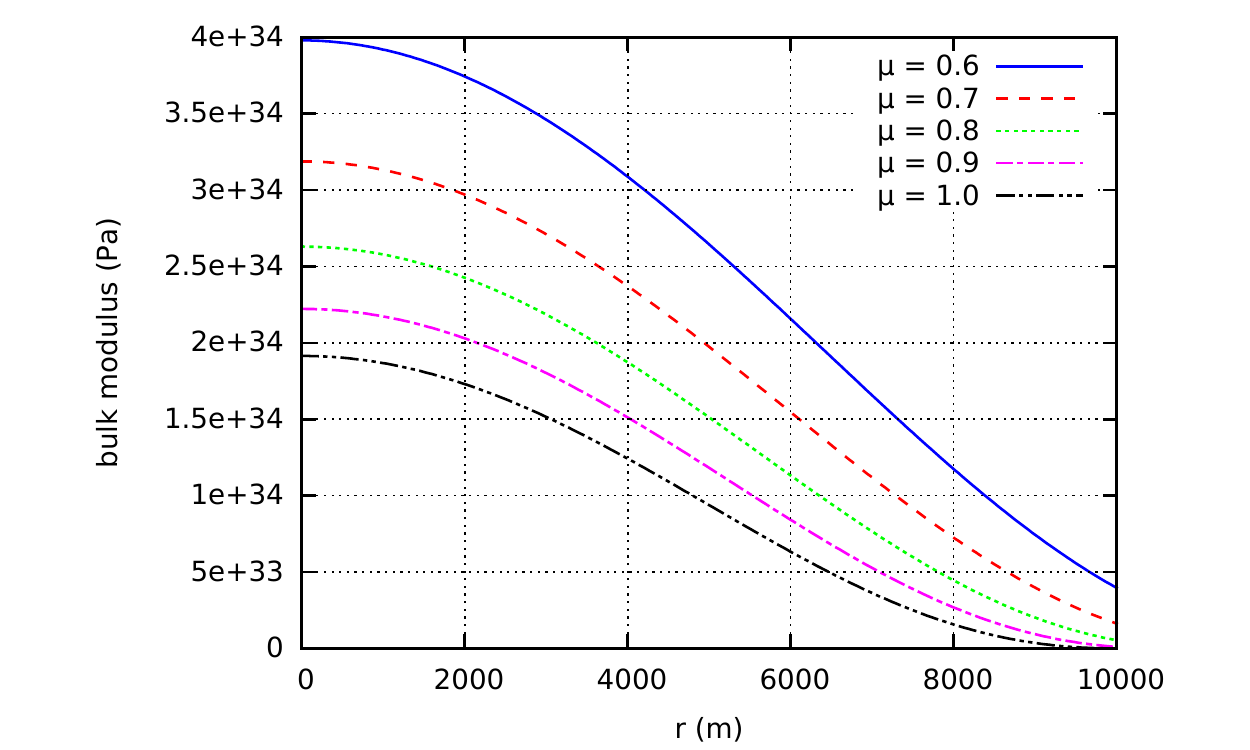}
\caption{Variation of bulk modulus with the radial coordinate
  inside the star. The parameter values are $\rho_{c}=\un{1\times
    10^{15} g \cdot cm^{-3}},$ $r_b = \un{1 \times 10^{6} cm}$ and $0.6 \leq \mu \leq 1.0$ .}
\label{fig:bulk}
\end{figure}

Similarly, the complicated expression for the pressure given 
by equations~\eqref{eq:PressureR},~\eqref{eq:C1solved} and~\eqref{eq:C2solved}
can also be plotted as a function of the radius.  Of importance here is the fact that
while the densities might not vanish at the boundary \(r_{b},\) the
pressure for all parameter values must do so according to the boundary
condition~\eqref{eq:BoundaryP}.  This is eminently clear in
Fig.~\ref{fig:pressure}, where we see the pressures associated with
the density curves shown in Fig.~\ref{fig:density}.  Similarly the
speed of sound and bulk modulus, all associated with the matter
content in the star are shown in
Figs.~\ref{fig:sound} and~\ref{fig:bulk} respectively.

\begin{figure}[h!]
\includegraphics[width=\linewidth]{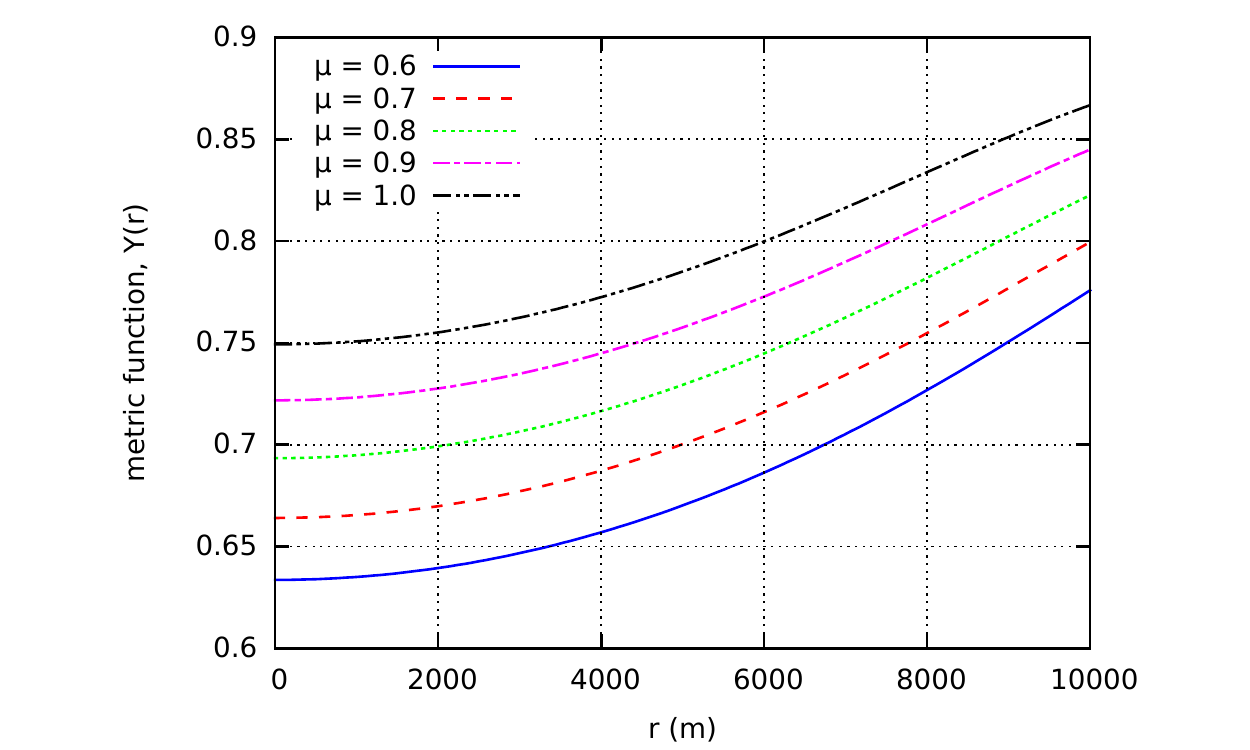}
\caption{Variation of \(Y\) metric variable with the radial coordinate
  inside the star. The parameter values are $\rho_{c}=\un{1\times
    10^{15} g \cdot cm^{-3}},$ $r_b = \un{1 \times 10^{6} cm}$ and $0.6 \leq \mu \leq 1.0$ .
  taking the various values shown in the legend.}
\label{fig:Ymetric}
\end{figure}

\begin{figure}[h!] 
\includegraphics[width=\linewidth]{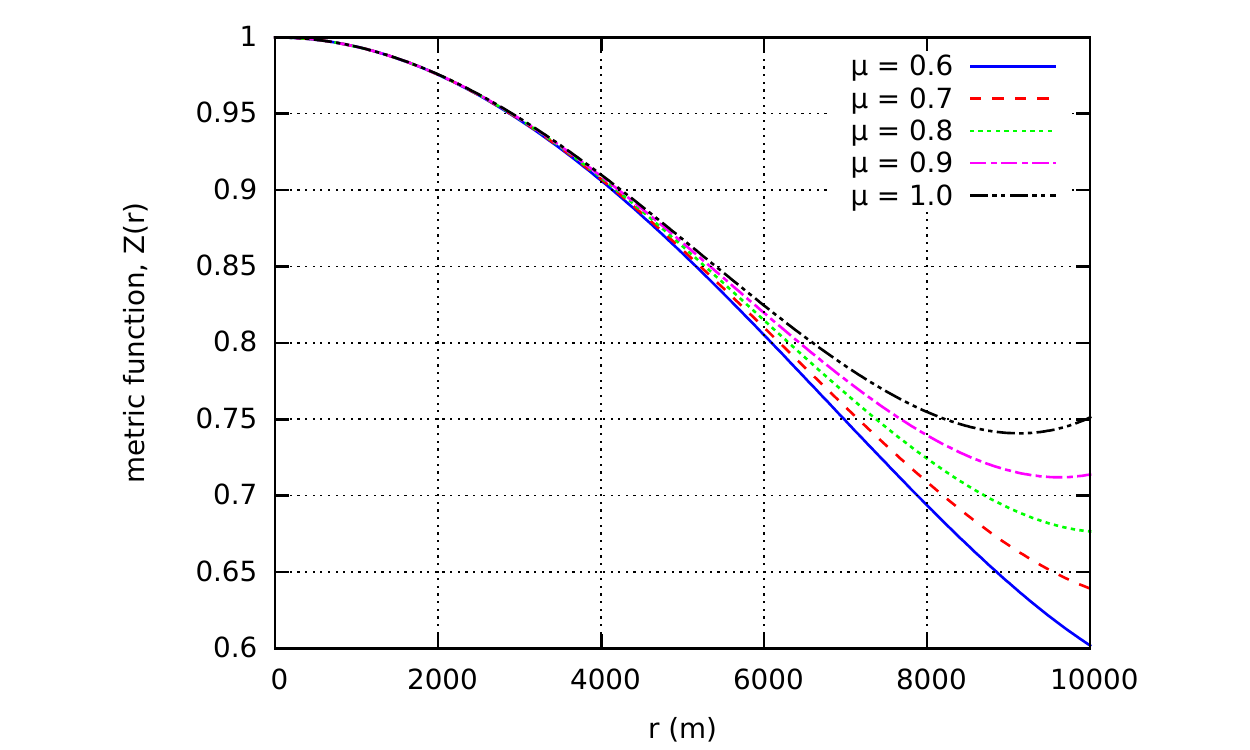}
\caption{Variation of \(Z\) metric variable with the radial coordinate
  inside the star. The parameter values are $\rho_{c}=\un{1\times
    10^{15} g \cdot cm^{-3}},$ $r_b = \un{1 \times 10^{6} m}$ and $0.6 \leq \mu \leq 1.0$ .}
\label{fig:Zmetric}
\end{figure}

\begin{figure}[h!]
\includegraphics[width=\linewidth]{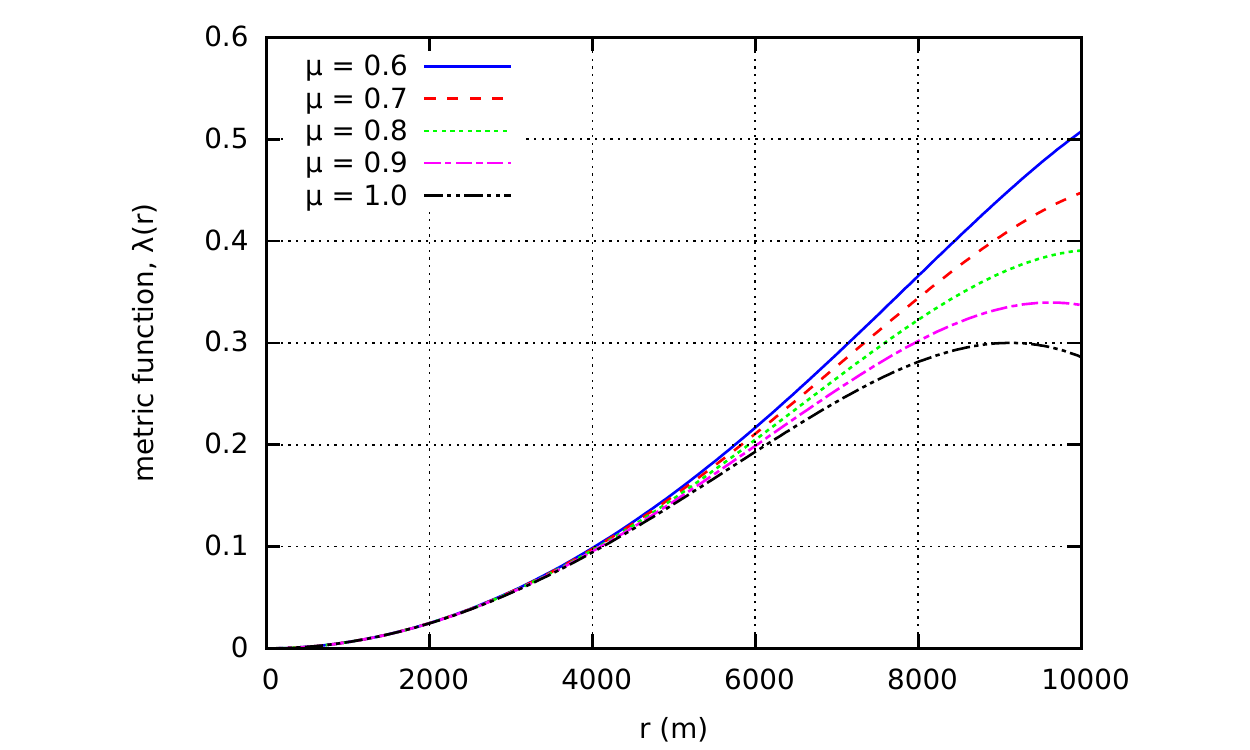}
\caption{Variation of \(\lambda\) metric variable with the radial coordinate
  inside the star. The parameter values are $\rho_{c}=\un{1\times
    10^{15} g \cdot m^{-3}}$ $r_b = \un{1 \times 10^{6} cm}$ and $0.6 \leq \mu \leq 1.0$ .}
\label{fig:LambdaMetric}
\end{figure}

\begin{figure}[h!]
\includegraphics[width=\linewidth]{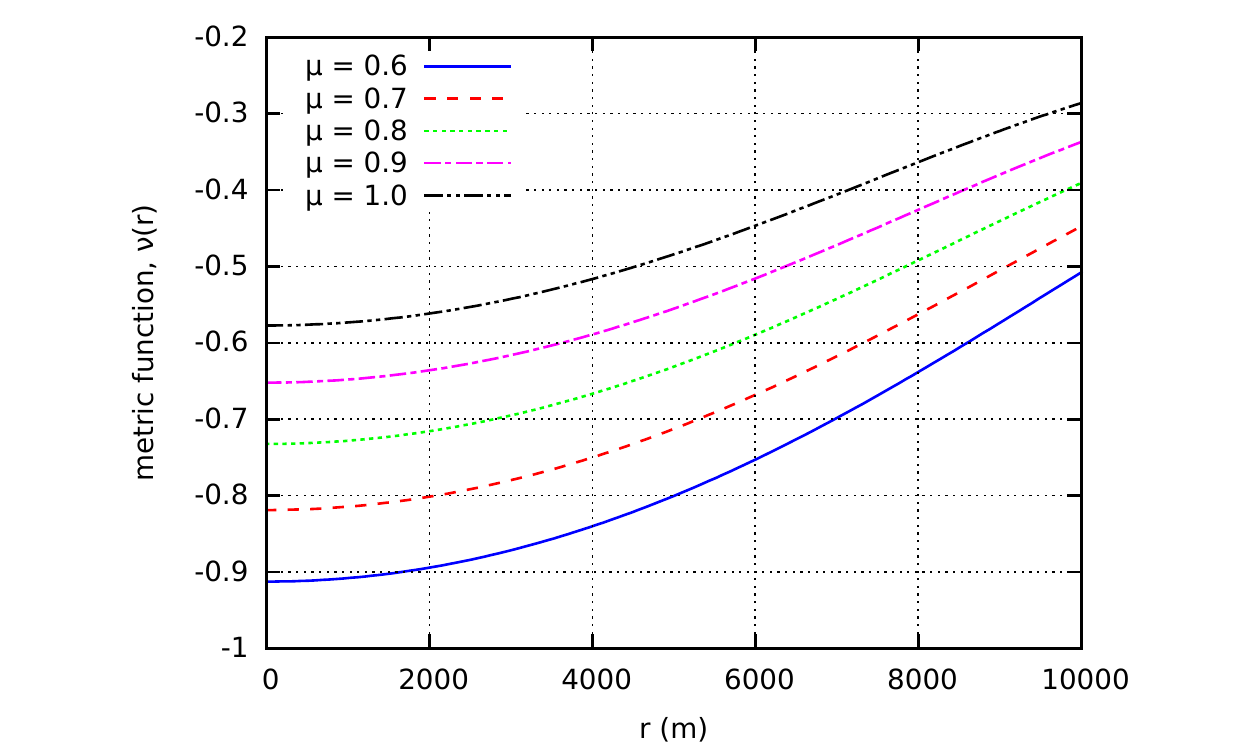}
\caption{Variation of \(\nu\) metric variable with the radial coordinate
  inside the star. The parameter values are $\rho_{c}=\un{1\times
    10^{15} g \cdot cm^{-3}}$ $r_b = \un{1 \times 10^{6} m}$ and $0.6 \leq \mu \leq 1.0$ .}
\label{fig:NuMetric}
\end{figure}

The functions $Z(r)$ and $Y(r)$ representing the solutions to the
differential equations (2) are given
in Figs.~\ref{fig:Zmetric} and~\ref{fig:Ymetric} respectively,
again for different values of the self-boundedness \(\mu.\) Equivalently
the metric coefficients in Schwarzschild form: the form
most often used in the literature for specifying static spherically
symmetric models can be obtained from $Y(r)$ and $Z(r)$.  For the sake of completeness, 
\(\lambda(r)\) is plotted in Fig.~\ref{fig:LambdaMetric} and \(\nu(r)\) is plotted in
Fig.~\ref{fig:NuMetric}.

\begin{figure}[h!]
  \centering
  \includegraphics[width=\linewidth]{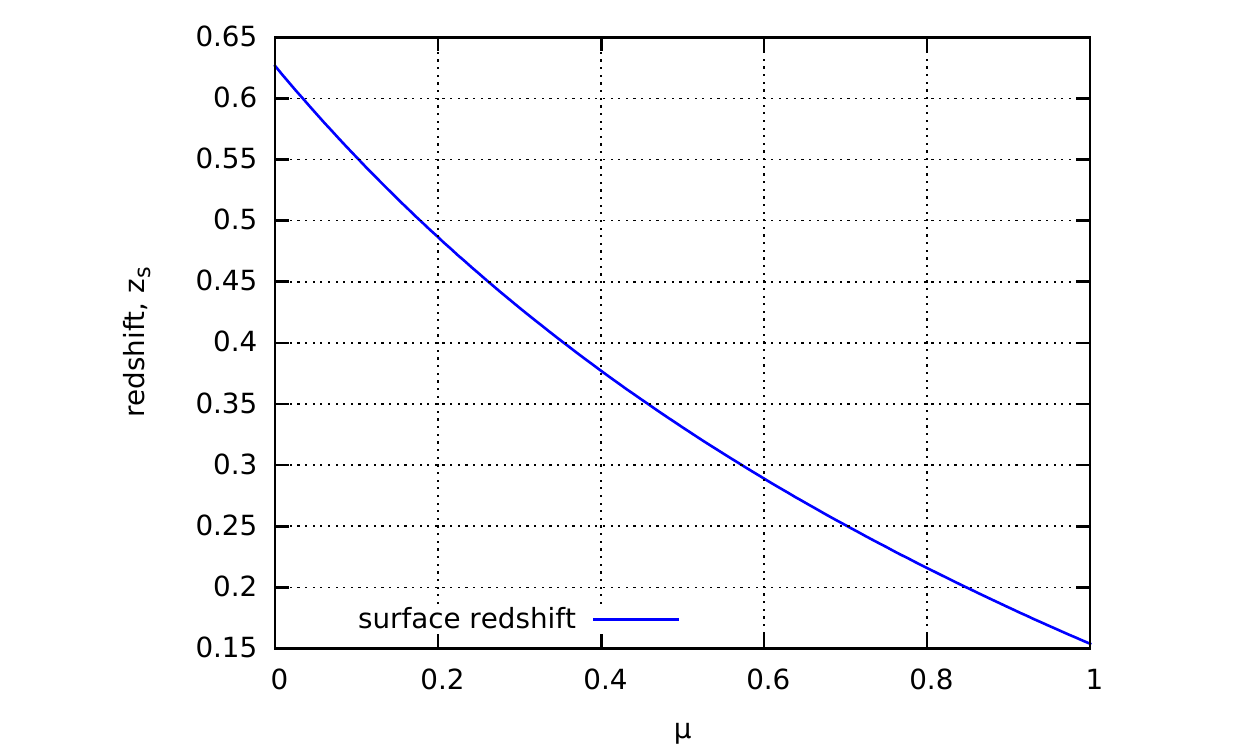}
  \caption{The redshift $z_{s}$ at the surface of the sphere for different values of $\mu.$}
  \label{fig:redshift}
\end{figure}

The redshift \(z_{s}\) of light emanating from a star as perceived by distant
observers is another quantity that potentially can be measured.  This
quantity can also be calculated in our model, from the relation \[z_{s} =
\left(1-\f{2m(r_{b})}{r_{b}} \right)^{-\f{1}{2}} - 1.\] The redshift value
at the surface of the star for different values of \(\mu\) is shown in
Fig.~\ref{fig:redshift}.

\section{\label{sec:eos}The equation of state and physical models}
A nice feature of the density assumption~\eqref{eq:Density} is that
it can be inverted to easily obtain \(r\) as a function of \(\rho.\)
This allows one to generate an equation of state (EOS) for this
solution.  The full equation of state is given below:
\[ p(\rho) = -\frac{1}{20 \pi h_1 h_2} \left\{ h_1 - h_2 \sqrt{-2 f_1
    \cot^{2} f_2} + 4 \pi h_1 h_2 \rho \right \},\] where \(f_1 (\rho)\) and
\(f_2(\rho)\) are functions of the density:
\[f_1(\rho) = 50 - 3\left(\f{h_{1}}{h_{2}}\right)^{2} -\f{4 \pi
  h_{1}^{2}}{h_{2}} \rho + 32 \pi^{2} h_{1}^{2} \rho^{2} \]
and
\[ f_2(\rho) = \frac{1}{2} \ln \left[ \frac{\sqrt {8 f_1 h_2} +
    h_1 - 16 \pi h_1 h_2 \rho }{20 h_2 C } \right]. \] The constants
\(h_1\) and \(h_2\) are determined by the central density and \(\mu,\)
as follows:
\[h_1 = r_b \sqrt{\frac{5}{2 \pi \rho_c \mu}} \qquad \text{and}
\qquad h_2 = \frac{3}{8 \pi \rho_c}, \] while the constant \(C\)
is expressible as a complicated function of the parameters only, in
terms of the auxiliary variables \(\sigma\) and \(\chi\):
\[
C = \left( 1 - \f{h_{1}}{4h_{2}}\right) \sqrt{\f{h_{1}(4h_{2}-h_{1})} {8r_{b}^{2} h_{2}-h_{1}^{2} + \chi} }
\exp{\left[\arctan\left(\f{\chi}{\sigma}\right)\right]},
\]
with
\begin{align*}
  \chi &= 4\sqrt{h_{2}(4h_{2}r_{b}^{4}-h_{1}^{2}r_{b}^{2} + h_{1}^{2}h_{2})},\\
  \sigma &= 16h_{2}r_{b}^{2}+8\pi \rho_{c} h_{1}^{2}h_{2}(1-\mu)-
  2h_{1}^{2}.
\end{align*}
It should be noted here that no assumption about the nature of matter, except
for the very general thermodynamic prescription of a perfect fluid, has
gone into this solution.  Everything else, and in particular the
equation of state, was obtained solely by virtue of the field equations
and the density profile~\eqref{eq:Density}.  With the equation of
state now given explicitly as \(p = p(\rho)\), it is a simple matter to find the derivative \(\rmd p / \rmd
\rho\) for the speed of pressure waves, and this yields precisely the
same function as the one found previously in
equation~\eqref{eq:SpeedSound}.

The expression for this class of EOS is somewhat complicated, but it
is not without physical interpretation, contrary to what
Tolman~\cite{Tol39} thought in 1939:
\begin{quote}
  The dependence of \(p\) on \(r\), with \(\rme^{-\lambda/2}\) and
  \(\rme^{-\nu}\) explicitly expressed in terms of \(r\), is so
  complicated that the solution is not a convenient one for physical
  considerations.
\end{quote}

By virtue of having a class of exact EOS, there is the possibility of
two separate interpretations for an EOS that arise from the analytic
expressions.  This
classification can be seen as a practical way of interpreting a class of
EOS that has four different parameters, not all independent of each
other.  Both \( \label{eq:EOS1} p(\rho; \I) \)
for \(\rho_{b} = \rho_{c}(1-\mu) \leq \rho \leq \rho_{c},\)
with the values of the elements of \( \I,\)
in particular \(\rho_{c},\)
fixed (henceforth called EOS1); and
\(\label{eq:EOS2} p(\rho=\rho_{c}; \I), \)
with the parameters of \(\I\)
varying between limits imposed by causality (EOS2), could be
candidates of the EOS.   

In the literature, both interpretations have been used, and sometimes
even interchanged.  However, each has a completely different content
in that the first interpretation expresses how the pressure of the
fluid changes in moving from the center of the star
\(r=0, \rho=\rho_{c},\)
to the boundary \(r=r_{b}, \rho_{b}=\rho_{c}(1-\mu),\)
while all the integration constants, and hence parameters \(\I,\)
are kept fixed.  Those seeking an interpretation of a unique EOS that
should be applicable to all neutron stars without exception would find
this interpretation sufficient.

The second interpretation, by contrast, looks closely at the fluid
material itself and how the pressure at a certain point in the star
changes as the density of the fluid at the center changes.  Given that
the central mass density of a compact star is inaccessible, this
interpretation is of interest to those who believe that the central
density should be a free parameter in a neutron star model.  This
would allow one to explore the possibilities that such a parameter
change has on the observable quantities of the stars.

At this point in the derivation, a causality condition has not been
imposed upon expressions of EOS1 or EOS2. Therefore, the class of EOS
obtained can take a wide range of parameter values, as long as the
metric functions and derived curvature tensors do not have
singularities.  The conditions leading to such singularities are
narrower than the causality criterion, and enforcing the latter
ensures that the parameter values do not cause singular behavior in
the solution, and hence the class of EOS.

\begin{figure}[h!]
\includegraphics[width=\linewidth]{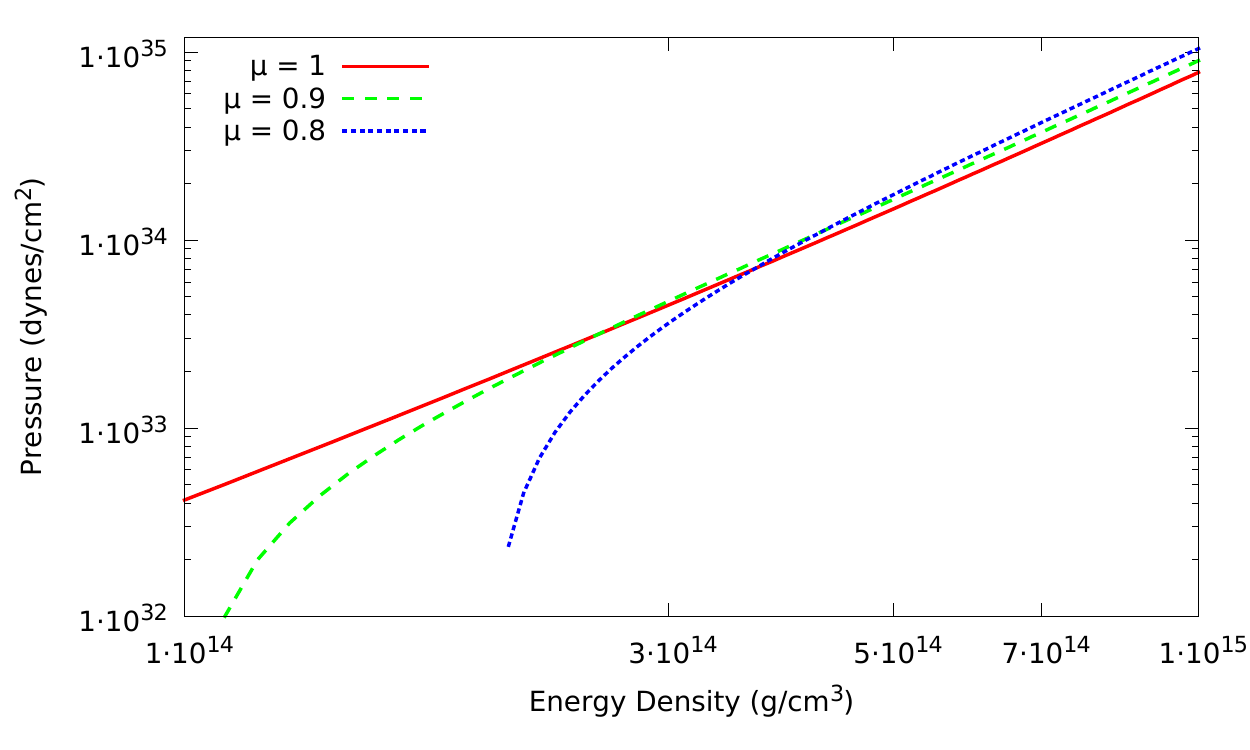}
\caption{\label{fig:loglog}Log-log plot of pressure
  versus density for neutron star models determined by different
  $\mu,$ but the same $\rho_{c}$ and $r_{b}.$ The densities and pressures
  are in cgs units, and the $\I$ is fixed by the following: $r_{b} =
  10^{6} \un{cm},$ $\rho_{c}=10^{15} \un{g \cdot cm^{-3}}.$ Since
  pressure is a decreasing function of distance from the center, large
  densities indicate points closer to the center of the star.}
\end{figure}

We first carry out an analysis of EOS1, and find that, to a high
degree of accuracy, the variation of \(p(\rho; \I),\)
with \(\rho,\)
and equivalently \(r,\)
is very close to that of a generalized polytrope of the general form
\(p = k \rho^{\gamma} - p_{0},\)
where $p_0$ is a pressure constant chosen such that $p$ vanishes when
$\rho = \rho_b$ at the boundary of the star and, as usual,$\gamma$ and
$k$ are the adiabatic index and the adiabatic constant, respectively.
This relation is very obvious from the shape of the curve in the
``natural'' \(\mu=1\)
case as is seen in the one curve in Fig.~\ref{fig:loglog}, and all the
curves in Figs.~\ref{fig:eosM} and~\ref{fig:eosR}. It is interesting
to note that the $\mu < 1$ cases all show a behavior similar to that
found for other self-bound EOSs. (See e.g.~Fig. 1 in the review by
\citeauthor*{LatPra01}~\cite{LatPra01}). Indeed, Fig.~\ref{fig:eosR}
even seems to suggest that varying the boundary radius \(r_{b}\)
changes the value of \(k\)
in the polytrope, and Fig.~\ref{fig:eosM} that varying \(\rho_{c}\)
changes the value of \(\gamma\) in the polytrope.

\begin{figure}[h!]
\includegraphics[width=\linewidth]{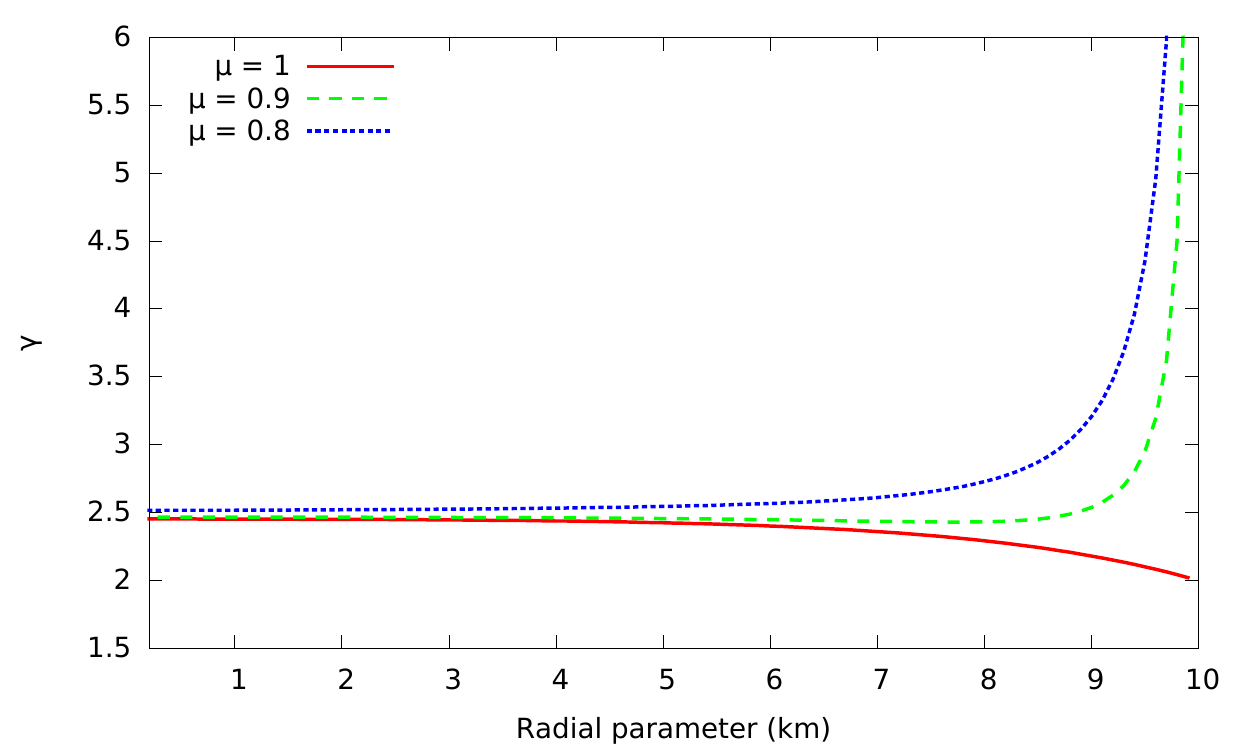}
\caption{\label{fig:gamma}The adiabatic index
  variation from the center to the boundary of the star for different
  values of the parameter $\mu.$ The other parameter values are the
  same as those in Fig.~\ref{fig:loglog}. }
\end{figure}

\begin{figure}[h!]
\includegraphics[width=\linewidth]{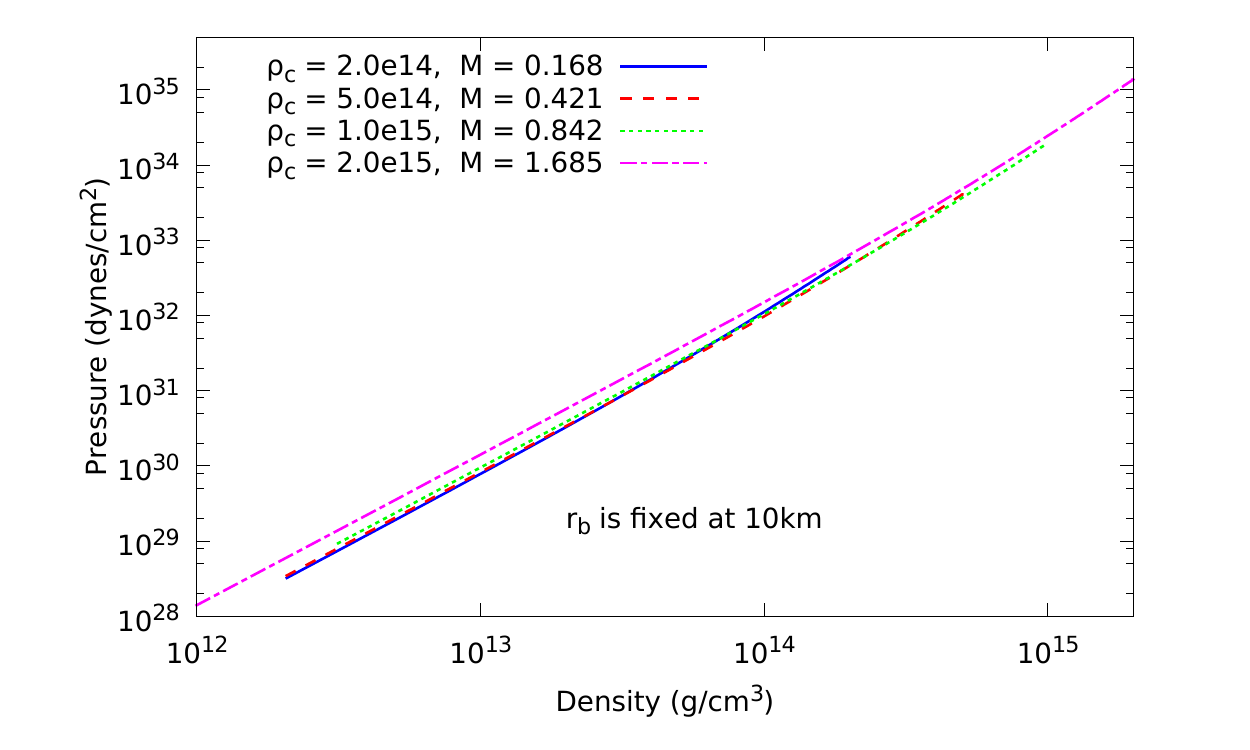}
\caption{\label{fig:eosM}The effect of changing the
  central density $\rho_{c}$ through 1 order of magnitude on the EOS
  with a boundary radius kept fixed at 10 km, for the natural EOS with
  $\mu=1$.  We notice that the slopes of the lines change by very
  little.  The legend also provides the corresponding masses associated
  with each parameter choice in solar mass units.}
\end{figure}

\begin{figure}[h!]
\includegraphics[width=\linewidth]{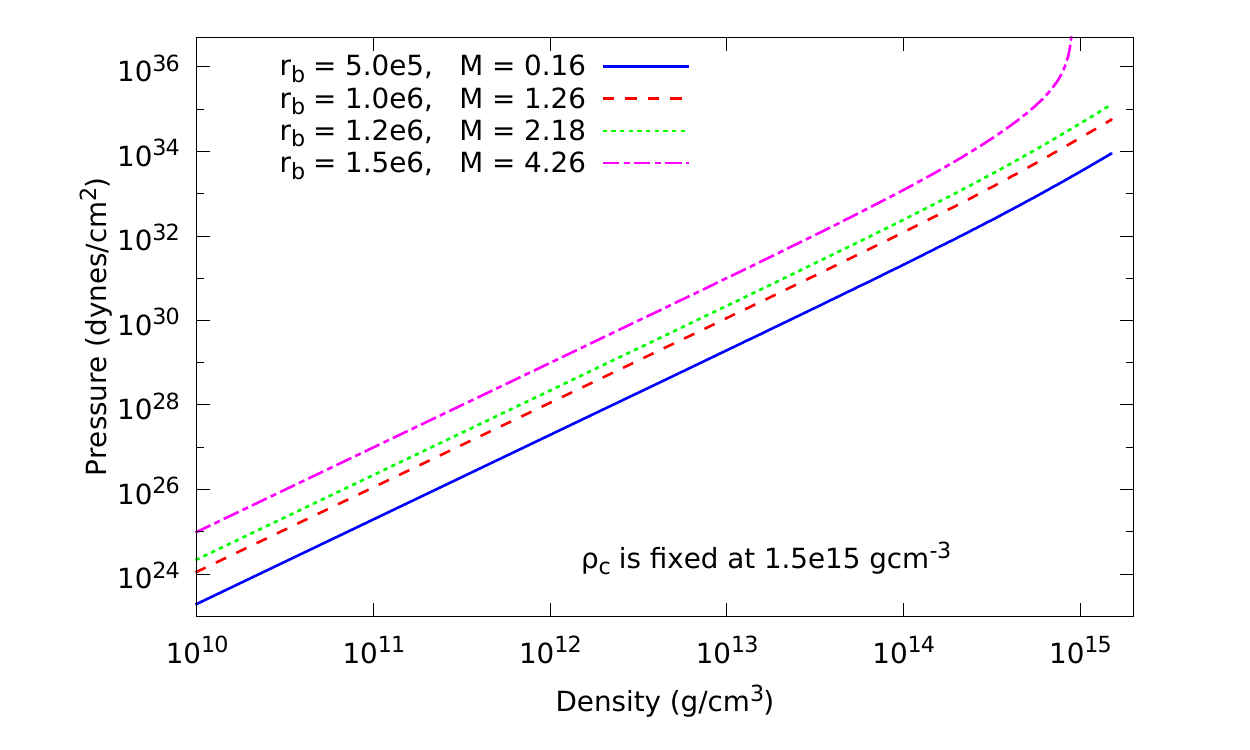}
\caption{\label{fig:eosR}The effect of changing the
  boundary radius $r_{b}$ through 1 order of magnitude, while the
  central density is kept fixed at
  $1.5 \times 10^{15}$ $\un{g \cdot cm^{-3}}.$ We notice that the lines
  remain mostly parallel, suggesting that only the $k$ value in the
  polytrope is changing, while the adiabatic index is remaining about
  the same. The masses of the corresponding stars are also provided in
  solar units in the legend.}
\end{figure}

Models employing polytropic perfect fluids use similar values for the
adiabatic index \(\gamma\) as what we find for a range of different
values of parameters \(\I.\) This is shown in Fig.~\ref{fig:gamma}
which treats \(\gamma\) as a continuous variable defined by \(\gamma =
\f{\rmd (\log p)}{\rmd (\log \rho)},\) and can be understood as the slope
of the previous log-log graph.

From this figure it becomes evident how both types of stars have an
interior structure well described by a polytrope with an adiabatic index close to
2.5.  The ``self-bound'' stars exhibit the existence of an envelope
consisting of material that is considerably stiffer than that found in
the interior.  Physically this is intuitive: for fixed \(\rho_{c}\)
and \(r_{b},\)
the self-bound stars will become more and more massive as \(\mu\)
decreases.  The increasing boundary density discontinuity requires a
stiffer exterior mass distribution to maintain the hydrostatic
equilibrium condition. 

A notable characteristic of the class of EOS is its uncanny ability to
distinguish between the different types of matter that make up the
natural and the self-bound stars. Since the mass
density~\eqref{eq:Density} is a monotonically decreasing function of
stellar radius, Fig.~\ref{fig:gamma} can be thought of as the
equivalent of a ``flipped'' and ``rescaled'' plot of the adiabatic
index as a function of density. For the case $\mu =1$,
Figs.~\ref{fig:loglog} and~\ref{fig:gamma} are consistent with a
number of hadronic EOS proposals.  For densities in the range of
1-10$\times 10^{14} \un{g \cdot cm^{-3}} $ the adiabatic index ranges
from 2.7 for the most dense nuclear material to 2.0 for the
lower-density material.  This type of behavior is found, for example,
in a model proposed by\citeauthor{Gle85}\cite{Gle85}, that consists of
a mixture of baryons.

For the self-bound models, where $\mu < 1,$ the Tolman EOS is
consistent with quark models using the MIT bag model.
Fig.~\ref{fig:loglog} is similar in nature to the strange quark models
(SQM1-3) shown in Fig.~1 in~\cite{LatPra01} while Fig.~\ref{fig:gamma}
has similarities to the work of \citeauthor*{CasMen10}, who analyze
the MIT bag model (see also the book
by~\citeauthor{HaePotYak07}~\cite{HaePotYak07}) and where it is found
that the EOS for quark matter stiffens significantly at low
densities~\cite{CasMen10}.  The adiabatic index of that material
reaches very high values (e.g.\ $\gamma > 7$ for densities less than
$10^{14}$ $\un{ g \cdot cm^{-3}}$).

To fully understand the nature of the EOS, the effect on the matter
resulting from changing the
other two parameters of \(\I\) can be investigated. Figure~\ref{fig:eosM} 
demonstrates how the EOS inside the star changes as
the central density \(\rho_{c}\) changes, and similarly Fig.~\ref{fig:eosR}
shows how the EOS varies with changes in the magnitude of the boundary radius
\(r_{b}\).

What these figures show is that the value of the free
parameters can change by up to an order of magnitude, yielding
drastically different masses, while maintaining the same general
polytropic behavior, \emph{independently} of the parameter choices.
At the highest densities, Fig.~\ref{fig:eosM} shows that the EOS are
nearly independent of the central density
parameter. Figure~\ref{fig:eosR} which demonstrates how the EOS changes
under changes in the boundary radius, indicates that the adiabatic
index $\gamma$ is fixed over a large range of densities, but as
expected the adiabatic constant $k$ is different for different stars.
The self-boundedness parameter \(\mu,\)
however, changes the character of this polytrope very much, as is clear
in Fig.~\ref{fig:gamma}, hinting that the same EOS can have a richer
structure than can be ultimately specified by the central density
alone.

This polytropic behavior is very satisfying, since we started by
trying to model a relativistic star from a Newtonian picture.  That a
class of EOS globally similar to the solutions of the Lane-Emden
equations becomes apparent when we extend the nonphysical
Schwarzschild interior to a more realistic density profile suggests
that Tolman~VII is at least as good as the Newtonian neutron stars,
however with relativity being taken into account.

Now turning to the second way to characterize the class of EOS,
concentrating on the behavior of the fluid material itself,
independent of the geometry of the star, we determine how different
physical quantities depend on the values of the central density
\(\rho_{c}.\) The total mass-energy is defined as
\begin{equation}
M = 4 \pi \int_{0}^{r_{b}} \bar{r}^{2} \rho(\bar{r}) \rmd \bar{r} = 
\f{4 \pi r_{b}^{3} \rho_{c} \left( 5-3\mu \right)}{15}.
\end{equation}
The mass is important, since it is the only directly and reliably
measurable quantity that can be obtained from neutron star
observations.  \citeauthor*{LatPra01}\cite{LatPra01, LatPra07,
  LatPra05} and others~\cite{Gle92,Gle96} have ruled out certain EOS2
based on mass and spin measurement of neutron stars.  The former have
also used Tolman~VII, to constrain other EOS2 based on nuclear
microphysics, and have even postulated that Tolman~VII could be used
as a guideline discriminating between viable and nonviable
EOS2~\cite{LatPra05}.  If this postulate is true, given that the
complete Tolman~VII EOS2 is known, the condition that the solution
must be causal can be applied, independent of measurements first, and
then compared with the previous works~\cite{Gle96,LatPra05}.

\begin{figure}[h!]
\includegraphics[width=\linewidth]{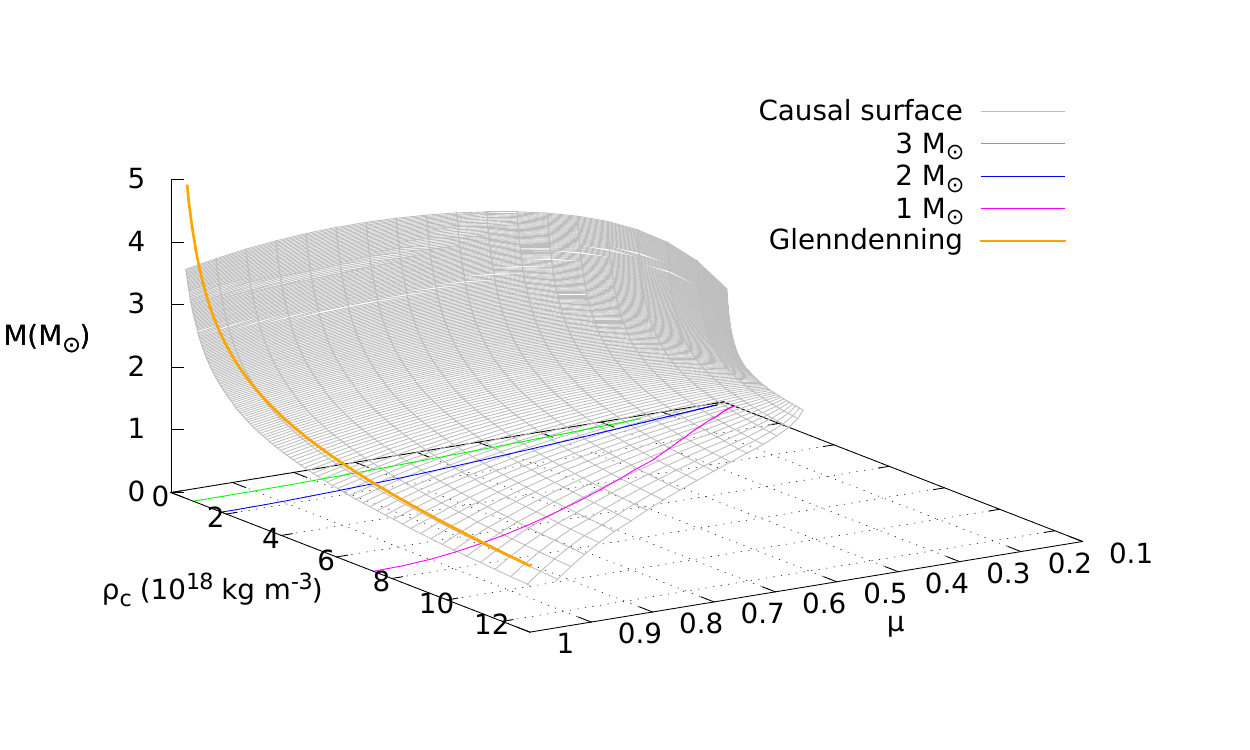}
\caption{\label{fig:phase}The mass of possible stars
  just obeying causality. The grey surface obeys the equation
  $v_{s}(r=0) =\un{c}.$ Every point below the surface is a possible
  realization of a star, and we can potentially read off the mass,
  central density, and $\mu$ value of that star.  The numbered lines
  represent stars with the same mass that are causal, i.e.\ they are
  projections of the causal surface onto the $\rho_{c}$-$\mu$ plane.
  Glendenning's~\cite{Gle96} curve is shown in orange and represents a
  limit in the natural case only, and according to our results is
  acausal, being above our surface.  The $\mu = 1$ plane's
  intersection with our graph is the graph given in Ref.~\cite{LatPra05},
  and here too our prediction is more restrictive.}
\end{figure}

This is done in Fig.~\ref{fig:phase}, where we superimpose the result
of Ref.~\cite{Gle96}, on our own analysis of the whole solution space
\(\I.\) The surface shown is that for the values at which the speed of
sound \(v_{s} = \left. \left(\sqrt{\rmd p / \rmd \rho} \right)
\right|_{r=0}\) at the center of the fluid sphere just reaches the
speed of light.  This is a sufficient condition for the solution to be
causal, since \(v_{s}\) is a monotonically decreasing function of \(r\)
in the sphere.  Any point located below this surface has coordinate
values for \(M, \rho_{c},\) and \(\mu\) that represent a valid
\emph{causal} solution to the Tolman~VII differential equations.  The
orange line is the previous result obtained by Glendenning~\cite{Gle96}
from rotational considerations.

Imposing causality to constrain the parameter space \(\I\)
is not a new idea.  However, having an explicit EOS allows one to
easily generate the causal surface shown in Fig.~\ref{fig:phase},
exactly without having to do any numerical gymnastics to find the
speed of sound.

\begin{figure}[h]
\includegraphics[width=\linewidth]{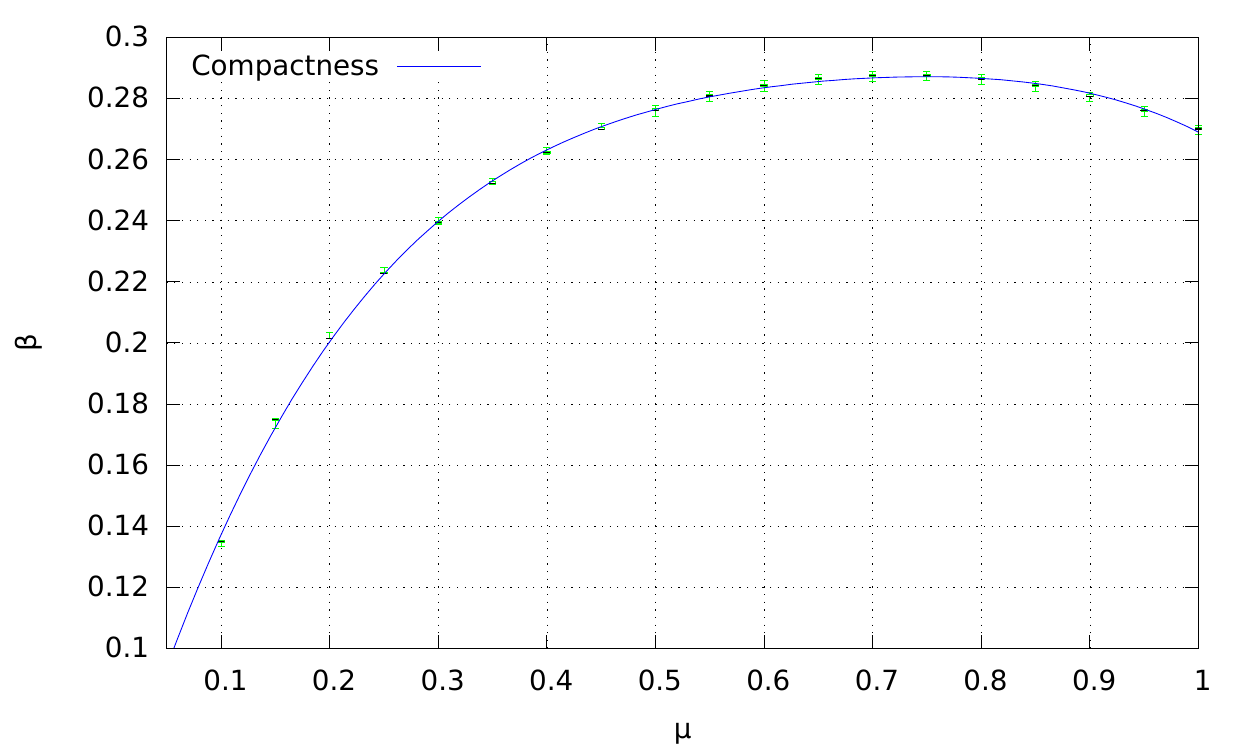}
\caption{\label{fig:beta}The compactness as a
  function of the self-boundedness parameter $\mu$.  This plot was
  generated by varying $r_{b}$ from 4 km to 20 km for fixed $\mu$ and
  finding $\rho_{c}$ and subsequently the compactness each time,
  such that the sound speed was causal at the center of the star.  The
  curve shown is a polynomial fit, and the box-and-whisker plots(very
  small in green) show the variation of $\beta$ for fixed $\mu$, but
  different $r_{b}.$ The very small whiskers justify the pertinence of
  $\beta$ as a useful measure in the analysis of the behavior of the
  model.}
\end{figure}
  
\begin{figure}[h]
\begin{center}
\includegraphics[width=\linewidth]{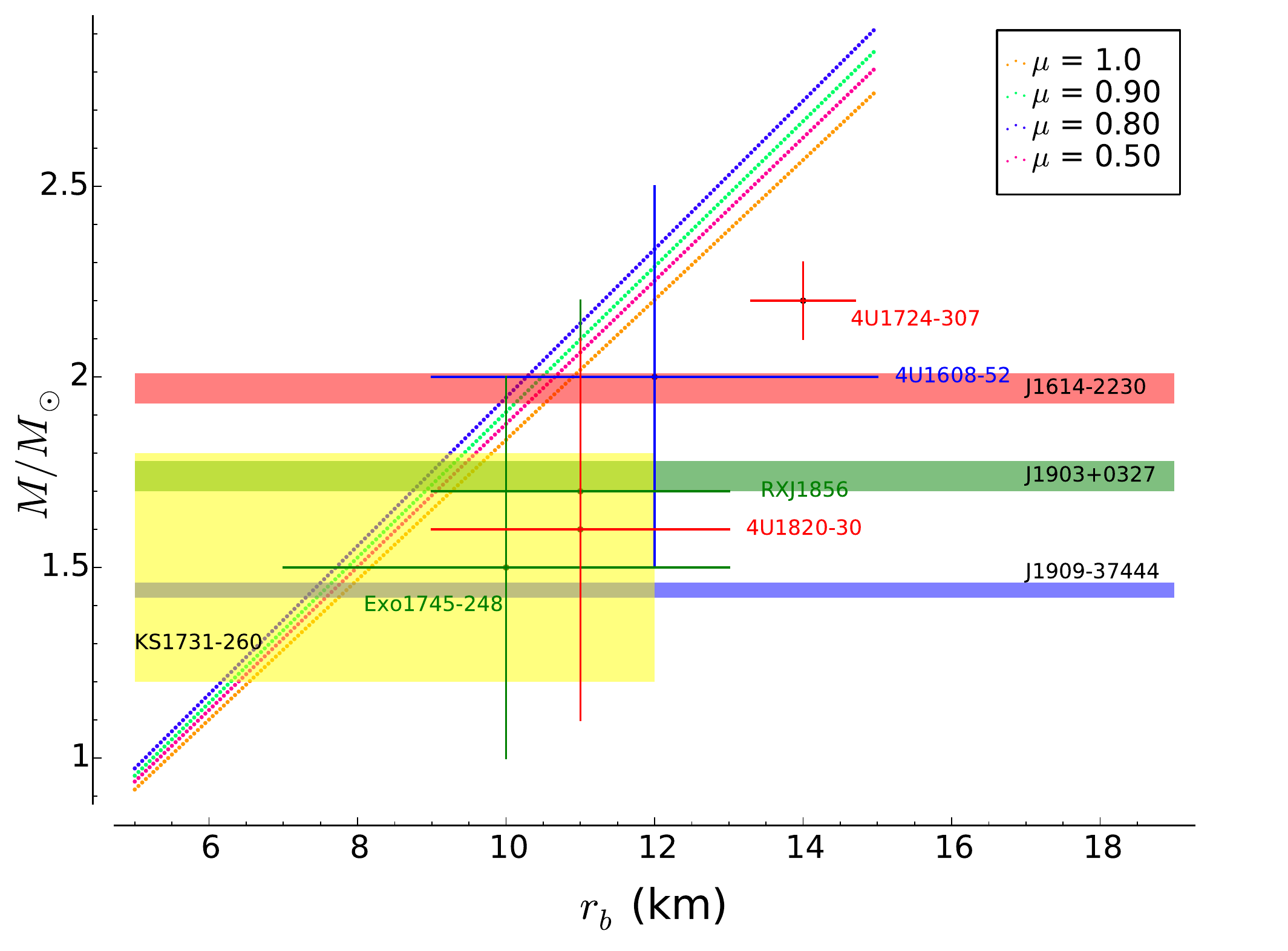}
\caption{\label{fig:expt}The mass \(M\)
  in solar units versus radius \(r_{b}\)
  in kilometers of a few stars for which these values have been
  measured. We use error bars to denote observational uncertainties,
  and colored bands in the case where only the mass is known.  The
  lines we show and the ones that are at the causality limit in the
  Tolman~VII model, and the whole area below those lines can be causal
  parameter choices for stars.}
\end{center}
\end{figure}

One method often used to distinguish between different EOS2 has been to
calculate the compactness ratio, given by
\(\label{eq:compactness} \beta = \f{\un{G} M}{\un{c^{2}}r_{b}}.\)
In the case of Tolman~VII the values of \(\beta\)
for a large range of parameter variations \(\I\)
are relatively constant.  This means that even though we might change
the value for \(\I\)
of the stars, the ones bordering on causality share very similar
compactness, albeit one that is lower than that previously thought
possible.  We show how this compactness \(\beta\)
varies with \(\mu\)
in Fig.~\ref{fig:beta}.  Clearly the variation of
$\beta$ is small (0.27 $< \beta <$ 0.29) over a large range of 
values for $\mu$ (0.45 $< \mu <$ 1.0).  A value for the maximal compactness of about
0.34 from rotational and causality criteria was obtained by Ref.~\cite{LatPra05}.  Our
analysis shows that \(\beta\)
is less than 0.3 for all possible stars, if Tolman~VII is a valid
physical model (at least as a limit) for stars.  Recently radius measurements
of a limited number of neutron stars have been
obtained~\cite{OzeGuvPsa08,*OzePsa09,*GuvWroCam10,*GuvOzeCab10,OzeGouGuv12,SulPou11,SteLatBro10}. These
are shown along with some other stars of known mass in
Fig.~\ref{fig:expt}.  We also superimpose a few of the limiting causal curves obtained for
different values of \(\mu\)
from Tolman~VII, to show that Tolman~VII is not ruled out by
observational results, even though it predicts lower compactness than
most nuclear models.  The dotted lines shown in the figure represent
the causal limits for different values of the self-boundedness
parameter $\mu$. 

For a fixed value of $\mu,$ causality requirements determine the
relationship between the central density and the boundary of the star.
A typical $M$-$R$ curve for a particular EOS2 would be determined by
fixing the parameters associated with a particular EOS2 and then
computing the mass as a function of the star's radius. All possible
curves of this nature lie to the right of the curves shown in
Fig.~\ref{fig:expt} and thus any curves in that region represent a
viable model for neutron star structure obtained from a Tolman~VII
solution.  This is one advantage of having a complete analytic
solution to the Einstein equations. Rather than integrate (using
numerical methods) the TOV equation for a specific set of EOS
parameters, an entire class of solutions are provided by the complete
analytic expressions. Physical constraints on the solution (such as
causality) then provide restrictions on that class of solutions.  What
the curves shown in this plot represent are the maximum possible cases
for the \((M,r_{b})\)
relation provided by the causal Tolman~VII EOS solutions.  Therefore,
the whole space to the right of those lines can potentially yield a
causal EOS and thus Fig.~\ref{fig:expt} shows that Tolman~VII is not
ruled out yet as a viable model for compact objects unless the radii
of some of the neutron stars with known masses are so small that they
would be represented by a point to the left of the curves.

The lines shown are on the edge of causality in the following way: the
speed of sound, a monotonically decreasing function of the radius, is
just equal to the speed of light at the center where \(r=0\)--
that is, the lines are the counterparts of those points that make up
the surface of Fig.~\ref{fig:phase}.  Since all observations of
compactness are bounded by the most extreme Tolman~VII model, we claim
that the solution may actually realized by compact stars in nature.

\section{\label{sec:conclusion}Conclusion}
A complete analysis of the Tolman~VII solution has been carried out,
and it was found that it is a physically valid solution with a huge
potential for modeling physical objects.  The class of EOS this
solution predicts has been found, and in certain regimes behaves very
much like a polytrope with an adiabatic index of 2.5, independently of
the choices of two parameters: the central density and the radius of the
star.  The third parameter, the self-boundedness, changes the
polytropic index drastically, particularly at the edge of the star, as
expected from a naive Newtonian approach to stellar structure.  That
this solution has a density profile that is very close to the
hand-picked, but thermodynamically motivated polytropes of Newtonian
stars, while still being a full relativistic model is a very good
reason to take its predictions of stellar structure seriously.

It is also interesting that the type of matter that produces the
different EOSs depends crucially on the value of $\mu$.  Hadronic
matter is obtained with $\mu = 1,$ while $\mu < 1$ stars would appear
to be made up of quark matter.

Using the EOS derived from Tolman~VII, we are able to compute the
speed of pressure waves, and imposing causality on the latter results
in a more restrictive limit on the maximum compactness of fluid
spheres allowable by classical general relativity.  This is possible
to do without the use of numerical computations because of the exact form
of the class of EOSs generated by the quadratic density falloff
assumption in Tolman~VII.

The solution is, moreover, stable under radial perturbations, since the
speed of these pressure waves is finite and monotonically decreasing
from the center outwards, thus satisfying the stability criterion
in Ref.~\cite{AbrHerNun07}.  If we believe as in Ref.~\cite{LatPra05} that
Tolman~VII is an upper limit on the possible energy density
\(\rho_{c},\)
for a given mass \(M,\)
some known models~\cite{LatPra07} which predict higher compactness
than Tolman~VII will have to be reconsidered, since they still
maintain a quadratic density profile to first order, and thus cannot
also be causal at these higher compactnesses.


\begin{thebibliography}{30}%
\makeatletter
\providecommand \@ifxundefined [1]{%
 \@ifx{#1\undefined}
}%
\providecommand \@ifnum [1]{%
 \ifnum #1\expandafter \@firstoftwo
 \else \expandafter \@secondoftwo
 \fi
}%
\providecommand \@ifx [1]{%
 \ifx #1\expandafter \@firstoftwo
 \else \expandafter \@secondoftwo
 \fi
}%
\providecommand \natexlab [1]{#1}%
\providecommand \enquote  [1]{``#1''}%
\providecommand \bibnamefont  [1]{#1}%
\providecommand \bibfnamefont [1]{#1}%
\providecommand \citenamefont [1]{#1}%
\providecommand \href@noop [0]{\@secondoftwo}%
\providecommand \href [0]{\begingroup \@sanitize@url \@href}%
\providecommand \@href[1]{\@@startlink{#1}\@@href}%
\providecommand \@@href[1]{\endgroup#1\@@endlink}%
\providecommand \@sanitize@url [0]{\catcode `\\12\catcode `\$12\catcode
  `\&12\catcode `\#12\catcode `\^12\catcode `\_12\catcode `\%12\relax}%
\providecommand \@@startlink[1]{}%
\providecommand \@@endlink[0]{}%
\providecommand \url  [0]{\begingroup\@sanitize@url \@url }%
\providecommand \@url [1]{\endgroup\@href {#1}{\urlprefix }}%
\providecommand \urlprefix  [0]{URL }%
\providecommand \Eprint [0]{\href }%
\providecommand \doibase [0]{http://dx.doi.org/}%
\providecommand \selectlanguage [0]{\@gobble}%
\providecommand \bibinfo  [0]{\@secondoftwo}%
\providecommand \bibfield  [0]{\@secondoftwo}%
\providecommand \translation [1]{[#1]}%
\providecommand \BibitemOpen [0]{}%
\providecommand \bibitemStop [0]{}%
\providecommand \bibitemNoStop [0]{.\EOS\space}%
\providecommand \EOS [0]{\spacefactor3000\relax}%
\providecommand \BibitemShut  [1]{\csname bibitem#1\endcsname}%
\let\auto@bib@innerbib\@empty
\bibitem [{\citenamefont {Kramer}\ \emph {et~al.}(1980)\citenamefont {Kramer},
  \citenamefont {Stephani}, \citenamefont {MacCallum},\ and\ \citenamefont
  {Herlt}}]{KraSteMac80}%
  \BibitemOpen
  \bibfield  {author} {\bibinfo {author} {\bibfnamefont {D.}~\bibnamefont
  {Kramer}}, \bibinfo {author} {\bibfnamefont {H.}~\bibnamefont {Stephani}},
  \bibinfo {author} {\bibfnamefont {M.~A.~H.}\ \bibnamefont {MacCallum}}, \
  and\ \bibinfo {author} {\bibfnamefont {E.}~\bibnamefont {Herlt}},\
  }\href@noop {} {\emph {\bibinfo {title} {Exact solutions of {Einstein's}
  field equations}}}\ (\bibinfo  {publisher} {Deutscher Verlag der
  Wissenschaften, Berlin, and Cambridge University Press},\ \bibinfo {address}
  {Cambridge},\ \bibinfo {year} {1980})\ pp.\ \bibinfo {pages} {1--425},\
  \bibinfo {note} {1.2}\BibitemShut {NoStop}%
\bibitem [{\citenamefont {{Delgaty}}\ and\ \citenamefont
  {{Lake}}(1998)}]{DelLak98}%
  \BibitemOpen
  \bibfield  {author} {\bibinfo {author} {\bibfnamefont {M.~S.~R.}\
  \bibnamefont {{Delgaty}}}\ and\ \bibinfo {author} {\bibfnamefont
  {K.}~\bibnamefont {{Lake}}},\ }\bibfield  {title} {\enquote {\bibinfo {title}
  {{Physical acceptability of isolated, static, spherically symmetric, perfect
  fluid solutions of Einstein's equations}},}\ }\href {\doibase
  10.1016/S0010-4655(98)00130-1} {\bibfield  {journal} {\bibinfo  {journal}
  {Computer Physics Communications}\ }\textbf {\bibinfo {volume} {115}},\
  \bibinfo {pages} {395--415} (\bibinfo {year} {1998})}\BibitemShut {NoStop}%
\bibitem [{\citenamefont {{Finch}}\ and\ \citenamefont
  {{Skea}}(1998)}]{FinSke98}%
  \BibitemOpen
  \bibfield  {author} {\bibinfo {author} {\bibfnamefont {M.~R.}\ \bibnamefont
  {{Finch}}}\ and\ \bibinfo {author} {\bibfnamefont {J.~E.~F}\ \bibnamefont
  {{Skea}}},\ }\href@noop {} {\enquote {\bibinfo {title} {A review of the
  relativistic static sphere},}\ } (\bibinfo {year} {1998}),\ \bibinfo {note}
  {unpublished, available at
  \url{www.dft.if.uerj.br/usuarios/JimSkea/papers/pfrev.ps}}\BibitemShut
  {NoStop}%
\bibitem [{\citenamefont {{Tolman}}(1939)}]{Tol39}%
  \BibitemOpen
  \bibfield  {author} {\bibinfo {author} {\bibfnamefont {R.~C.}\ \bibnamefont
  {{Tolman}}},\ }\bibfield  {title} {\enquote {\bibinfo {title} {{Static
  Solutions of Einstein's Field Equations for Spheres of Fluid}},}\ }\href
  {\doibase 10.1103/PhysRev.55.364} {\bibfield  {journal} {\bibinfo  {journal}
  {Physical Review}\ }\textbf {\bibinfo {volume} {55}},\ \bibinfo {pages}
  {364--373} (\bibinfo {year} {1939})}\BibitemShut {NoStop}%
\bibitem [{\citenamefont {{Kinnersley}}(1975)}]{Kin75}%
  \BibitemOpen
  \bibfield  {author} {\bibinfo {author} {\bibfnamefont {W.}~\bibnamefont
  {{Kinnersley}}},\ }\bibfield  {title} {\enquote {\bibinfo {title} {{Recent
  Progress in Exact Solutions}},}\ }in\ \href@noop {} {\emph {\bibinfo
  {booktitle} {Relativity and Gravitation}}},\ \bibinfo {editor} {edited by\
  \bibinfo {editor} {\bibfnamefont {G.}~\bibnamefont {{Shaviv}}}\ and\ \bibinfo
  {editor} {\bibfnamefont {J.}~\bibnamefont {{Rosen}}}}\ (\bibinfo {year}
  {1975})\ p.\ \bibinfo {pages} {109}\BibitemShut {NoStop}%
\bibitem [{\citenamefont {Durgapal}\ and\ \citenamefont
  {Gehlot}(1971)}]{DurGeh71}%
  \BibitemOpen
  \bibfield  {author} {\bibinfo {author} {\bibfnamefont {M.~C.}\ \bibnamefont
  {Durgapal}}\ and\ \bibinfo {author} {\bibfnamefont {G.~L.}\ \bibnamefont
  {Gehlot}},\ }\bibfield  {title} {\enquote {\bibinfo {title} {Spheres with
  varying density in general relativity},}\ }\href
  {http://stacks.iop.org/0022-3689/4/749} {\bibfield  {journal} {\bibinfo
  {journal} {Journal of Physics A: General Physics}\ }\textbf {\bibinfo
  {volume} {4}},\ \bibinfo {pages} {749--755} (\bibinfo {year}
  {1971})}\BibitemShut {NoStop}%
\bibitem [{\citenamefont {{Durgapal}}\ and\ \citenamefont
  {{Rawat}}(1980)}]{DurRaw79}%
  \BibitemOpen
  \bibfield  {author} {\bibinfo {author} {\bibfnamefont {M.~C.}\ \bibnamefont
  {{Durgapal}}}\ and\ \bibinfo {author} {\bibfnamefont {P.~S.}\ \bibnamefont
  {{Rawat}}},\ }\bibfield  {title} {\enquote {\bibinfo {title} {{Non-rigid
  massive spheres in general relativity}},}\ }\href@noop {} {\bibfield
  {journal} {\bibinfo  {journal} {Mon. Not. R. Astron. Soc.}\ }\textbf
  {\bibinfo {volume} {192}},\ \bibinfo {pages} {659--662} (\bibinfo {year}
  {1980})}\BibitemShut {NoStop}%
\bibitem [{\citenamefont {Mehra}(1966)}]{Meh66}%
  \BibitemOpen
  \bibfield  {author} {\bibinfo {author} {\bibfnamefont {A.~L.}\ \bibnamefont
  {Mehra}},\ }\bibfield  {title} {\enquote {\bibinfo {title} {Radially
  symmetric distribution of matter},}\ }\href {\doibase
  10.1017/S1446788700004730} {\bibfield  {journal} {\bibinfo  {journal}
  {Journal of the Australian Mathematical Society}\ }\textbf {\bibinfo {volume}
  {6}},\ \bibinfo {pages} {153--156} (\bibinfo {year} {1966})}\BibitemShut
  {NoStop}%
\bibitem [{\citenamefont {{Neary}}\ and\ \citenamefont
  {{Lake}}(2001)}]{NeaLak01}%
  \BibitemOpen
  \bibfield  {author} {\bibinfo {author} {\bibfnamefont {N.}~\bibnamefont
  {{Neary}}}\ and\ \bibinfo {author} {\bibfnamefont {K.}~\bibnamefont
  {{Lake}}},\ }\bibfield  {title} {\enquote {\bibinfo {title} {{r-modes in the
  Tolman VII solution}},}\ }\href@noop {} {\bibfield  {journal} {\bibinfo
  {journal} {ArXiv General Relativity and Quantum Cosmology e-prints}\ }
  (\bibinfo {year} {2001})},\ \Eprint {http://arxiv.org/abs/gr-qc/0106056}
  {gr-qc/0106056} \BibitemShut {NoStop}%
\bibitem [{\citenamefont {Neary}\ \emph {et~al.}(2001)\citenamefont {Neary},
  \citenamefont {Ishak},\ and\ \citenamefont {Lake}}]{NeaIshLak01}%
  \BibitemOpen
  \bibfield  {author} {\bibinfo {author} {\bibfnamefont {Nicholas}\
  \bibnamefont {Neary}}, \bibinfo {author} {\bibfnamefont {Mustapha}\
  \bibnamefont {Ishak}}, \ and\ \bibinfo {author} {\bibfnamefont {Kayll}\
  \bibnamefont {Lake}},\ }\bibfield  {title} {\enquote {\bibinfo {title} {The
  {Tolman VII} solution, trapped null orbits and w - modes},}\ }\href
  {http://www.citebase.org/abstract?id=oai:arXiv.org:gr-qc/0104002} {\bibfield
  {journal} {\bibinfo  {journal} {Physical Review D}\ }\textbf {\bibinfo
  {volume} {64}},\ \bibinfo {pages} {084001} (\bibinfo {year}
  {2001})}\BibitemShut {NoStop}%
\bibitem [{\citenamefont {Horedt}(2004)}]{Hor04}%
  \BibitemOpen
  \bibfield  {author} {\bibinfo {author} {\bibfnamefont {G.P.}\ \bibnamefont
  {Horedt}},\ }\href@noop {} {\emph {\bibinfo {title} {Polytropes: Applications
  in Astrophysics and Related Fields}}},\ Astrophysics and Space Science
  Library\ (\bibinfo  {publisher} {Springer Netherlands},\ \bibinfo {year}
  {2004})\BibitemShut {NoStop}%
\bibitem [{\citenamefont {Buchdahl}(1959)}]{Buc59}%
  \BibitemOpen
  \bibfield  {author} {\bibinfo {author} {\bibfnamefont {H.~A.}\ \bibnamefont
  {Buchdahl}},\ }\bibfield  {title} {\enquote {\bibinfo {title} {General
  relativistic fluid spheres},}\ }\href {\doibase 10.1103/PhysRev.116.1027}
  {\bibfield  {journal} {\bibinfo  {journal} {Phys. Rev.}\ }\textbf {\bibinfo
  {volume} {116}},\ \bibinfo {pages} {1027--1034} (\bibinfo {year}
  {1959})}\BibitemShut {NoStop}%
\bibitem [{\citenamefont {{Lattimer}}\ and\ \citenamefont
  {{Prakash}}(2001)}]{LatPra01}%
  \BibitemOpen
  \bibfield  {author} {\bibinfo {author} {\bibfnamefont {J.~M.}\ \bibnamefont
  {{Lattimer}}}\ and\ \bibinfo {author} {\bibfnamefont {M.}~\bibnamefont
  {{Prakash}}},\ }\bibfield  {title} {\enquote {\bibinfo {title} {{Neutron Star
  Structure and the Equation of State}},}\ }\href {\doibase 10.1086/319702}
  {\bibfield  {journal} {\bibinfo  {journal} {The Astrophysical Journal}\
  }\textbf {\bibinfo {volume} {550}},\ \bibinfo {pages} {426--442} (\bibinfo
  {year} {2001})},\ \bibinfo {note} {in particular Fig. 5 shows the quadradic
  fall-off behaviour of the density}\BibitemShut {NoStop}%
\bibitem [{\citenamefont {Ivanov}(2002)}]{Iva02}%
  \BibitemOpen
  \bibfield  {author} {\bibinfo {author} {\bibfnamefont {B.~V.}\ \bibnamefont
  {Ivanov}},\ }\bibfield  {title} {\enquote {\bibinfo {title} {Static charged
  perfect fluid spheres in general relativity},}\ }\href
  {http://www.citebase.org/abstract?id=oai:arXiv.org:gr-qc/0203070} {\bibfield
  {journal} {\bibinfo  {journal} {Physical Review D}\ }\textbf {\bibinfo
  {volume} {65}},\ \bibinfo {pages} {104001} (\bibinfo {year}
  {2002})}\BibitemShut {NoStop}%
\bibitem [{\citenamefont {Synge}(1960)}]{Syn60}%
  \BibitemOpen
  \bibfield  {author} {\bibinfo {author} {\bibfnamefont {J.L.}\ \bibnamefont
  {Synge}},\ }\href@noop {} {\emph {\bibinfo {title} {Relativity: the general
  theory}}},\ Series in physics\ (\bibinfo  {publisher} {North-Holland Pub.
  Co.},\ \bibinfo {year} {1960})\BibitemShut {NoStop}%
\bibitem [{\citenamefont {{Glendenning}}(1985)}]{Gle85}%
  \BibitemOpen
  \bibfield  {author} {\bibinfo {author} {\bibfnamefont {N.~K.}\ \bibnamefont
  {{Glendenning}}},\ }\bibfield  {title} {\enquote {\bibinfo {title} {{Neutron
  stars are giant hypernuclei?}}}\ }\href {\doibase 10.1086/163253} {\bibfield
  {journal} {\bibinfo  {journal} {\apj}\ }\textbf {\bibinfo {volume} {293}},\
  \bibinfo {pages} {470--493} (\bibinfo {year} {1985})}\BibitemShut {NoStop}%
\bibitem [{\citenamefont {Casali}\ and\ \citenamefont
  {Menezes}(2010)}]{CasMen10}%
  \BibitemOpen
  \bibfield  {author} {\bibinfo {author} {\bibfnamefont {R.~H.}\ \bibnamefont
  {Casali}}\ and\ \bibinfo {author} {\bibfnamefont {D.~P.}\ \bibnamefont
  {Menezes}},\ }\bibfield  {title} {\enquote {\bibinfo
  {title} {{Adiabatic index of hot and cold compact objects}},}\ }\href@noop
  {} {\bibfield  {journal} {\bibinfo  {journal} {{Brazilian Journal of
  Physics}}\ }\textbf {\bibinfo {volume} {40}},\ \bibinfo {pages} {166 -- 171}
  (\bibinfo {year} {2010})}\BibitemShut {NoStop}%
\bibitem [{\citenamefont {{Haensel}}\ \emph {et~al.}(2007)\citenamefont
  {{Haensel}}, \citenamefont {{Potekin}},\ and\ \citenamefont
  {{Yakovlev}}}]{HaePotYak07}%
  \BibitemOpen
  \bibfield  {author} {\bibinfo {author} {\bibfnamefont {P.}~\bibnamefont
  {{Haensel}}}, \bibinfo {author} {\bibfnamefont {A.Y}\ \bibnamefont
  {{Potekin}}}, \ and\ \bibinfo {author} {\bibfnamefont {D.G}\ \bibnamefont
  {{Yakovlev}}},\ }\href@noop {} {\emph {\bibinfo {title} {Neutron~Stars~1 :
  Equation of State and Structure}}},\ Vol.~\bibinfo {volume} {1}\ (\bibinfo
  {publisher} {Springer},\ \bibinfo {year} {2007})\BibitemShut {NoStop}%
\bibitem [{\citenamefont {{Lattimer}}\ and\ \citenamefont
  {{Prakash}}(2007)}]{LatPra07}%
  \BibitemOpen
  \bibfield  {author} {\bibinfo {author} {\bibfnamefont {J.~M.}\ \bibnamefont
  {{Lattimer}}}\ and\ \bibinfo {author} {\bibfnamefont {M.}~\bibnamefont
  {{Prakash}}},\ }\bibfield  {title} {\enquote {\bibinfo {title} {{Neutron star
  observations: Prognosis for equation of state constraints}},}\ }\href
  {\doibase 10.1016/j.physrep.2007.02.003} {\bibfield  {journal} {\bibinfo
  {journal} {Phys. Rep.}\ }\textbf {\bibinfo {volume} {442}},\ \bibinfo {pages}
  {109--165} (\bibinfo {year} {2007})}\BibitemShut {NoStop}%
\bibitem [{\citenamefont {Lattimer}\ and\ \citenamefont
  {Prakash}(2005)}]{LatPra05}%
  \BibitemOpen
  \bibfield  {author} {\bibinfo {author} {\bibfnamefont {James~M.}\
  \bibnamefont {Lattimer}}\ and\ \bibinfo {author} {\bibfnamefont {Madappa}\
  \bibnamefont {Prakash}},\ }\bibfield  {title} {\enquote {\bibinfo {title}
  {Ultimate energy density of observable cold baryonic matter},}\ }\href
  {\doibase 10.1103/PhysRevLett.94.111101} {\bibfield  {journal} {\bibinfo
  {journal} {Phys. Rev. Lett.}\ }\textbf {\bibinfo {volume} {94}},\ \bibinfo
  {pages} {111101} (\bibinfo {year} {2005})}\BibitemShut {NoStop}%
\bibitem [{\citenamefont {Glendenning}(1992)}]{Gle92}%
  \BibitemOpen
  \bibfield  {author} {\bibinfo {author} {\bibfnamefont {Norman~K.}\
  \bibnamefont {Glendenning}},\ }\bibfield  {title} {\enquote {\bibinfo {title}
  {First-order phase transitions with more than one conserved charge:
  Consequences for neutron stars},}\ }\href {\doibase 10.1103/PhysRevD.46.1274}
  {\bibfield  {journal} {\bibinfo  {journal} {Phys. Rev. D}\ }\textbf {\bibinfo
  {volume} {46}},\ \bibinfo {pages} {1274--1287} (\bibinfo {year}
  {1992})}\BibitemShut {NoStop}%
\bibitem [{\citenamefont {{Glendenning}}(1996)}]{Gle96}%
  \BibitemOpen
  \bibfield  {author} {\bibinfo {author} {\bibfnamefont {N.}~\bibnamefont
  {{Glendenning}}},\ }\href@noop {} {\emph {\bibinfo {title} {{Compact Stars.
  Nuclear Physics, Particle Physics and General Relativity.}}}}\ (\bibinfo
  {publisher} {Springer-Verlag New York},\ \bibinfo {year} {1996})\BibitemShut
  {NoStop}%
\bibitem [{\citenamefont {\"Ozel}\ \emph {et~al.}(2009)\citenamefont {\"Ozel},
  \citenamefont {G\"uver},\ and\ \citenamefont {Psaltis}}]{OzeGuvPsa08}%
  \BibitemOpen
  \bibfield  {author} {\bibinfo {author} {\bibfnamefont {Feryal}\ \bibnamefont
  {\"Ozel}}, \bibinfo {author} {\bibfnamefont {Tolga}\ \bibnamefont {G\"uver}},
  \ and\ \bibinfo {author} {\bibfnamefont {Dimitrios}\ \bibnamefont
  {Psaltis}},\ }\bibfield  {title} {\enquote {\bibinfo {title} {The mass and
  radius of the neutron star in exo 1745–248},}\ }\href
  {http://stacks.iop.org/0004-637X/693/i=2/a=1775} {\bibfield  {journal}
  {\bibinfo  {journal} {The Astrophysical Journal}\ }\textbf {\bibinfo {volume}
  {693}},\ \bibinfo {pages} {1775} (\bibinfo {year} {2009})}\BibitemShut
  {NoStop}%
\bibitem [{\citenamefont {\"Ozel}\ and\ \citenamefont
  {Psaltis}(2009)}]{OzePsa09}%
  \BibitemOpen
  \bibfield  {author} {\bibinfo {author} {\bibfnamefont {Feryal}\ \bibnamefont
  {\"Ozel}}\ and\ \bibinfo {author} {\bibfnamefont {Dimitrios}\ \bibnamefont
  {Psaltis}},\ }\bibfield  {title} {\enquote {\bibinfo {title} {Reconstructing
  the neutron-star equation of state from astrophysical measurements},}\ }\href
  {\doibase 10.1103/PhysRevD.80.103003} {\bibfield  {journal} {\bibinfo
  {journal} {Phys. Rev. D}\ }\textbf {\bibinfo {volume} {80}},\ \bibinfo
  {pages} {103003} (\bibinfo {year} {2009})}\BibitemShut {NoStop}%
\bibitem [{\citenamefont {G\"uver}\ \emph {et~al.}(2010)\citenamefont
  {G\"uver}, \citenamefont {Wroblewski}, \citenamefont {Camarota},\ and\
  \citenamefont {\"Ozel}}]{GuvWroCam10}%
  \BibitemOpen
  \bibfield  {author} {\bibinfo {author} {\bibfnamefont {Tolga}\ \bibnamefont
  {G\"uver}}, \bibinfo {author} {\bibfnamefont {Patricia}\ \bibnamefont
  {Wroblewski}}, \bibinfo {author} {\bibfnamefont {Larry}\ \bibnamefont
  {Camarota}}, \ and\ \bibinfo {author} {\bibfnamefont {Feryal}\ \bibnamefont
  {\"Ozel}},\ }\bibfield  {title} {\enquote {\bibinfo {title} {The mass and
  radius of the neutron star in 4u 1820–30},}\ }\href
  {http://stacks.iop.org/0004-637X/719/i=2/a=1807} {\bibfield  {journal}
  {\bibinfo  {journal} {The Astrophysical Journal}\ }\textbf {\bibinfo {volume}
  {719}},\ \bibinfo {pages} {1807} (\bibinfo {year} {2010})}\BibitemShut
  {NoStop}%
\bibitem [{\citenamefont {{G{\"u}ver}}\ \emph {et~al.}(2010)\citenamefont
  {{G{\"u}ver}}, \citenamefont {{{\"O}zel}}, \citenamefont {{Cabrera-Lavers}},\
  and\ \citenamefont {{Wroblewski}}}]{GuvOzeCab10}%
  \BibitemOpen
  \bibfield  {author} {\bibinfo {author} {\bibfnamefont {T.}~\bibnamefont
  {{G{\"u}ver}}}, \bibinfo {author} {\bibfnamefont {F.}~\bibnamefont
  {{{\"O}zel}}}, \bibinfo {author} {\bibfnamefont {A.}~\bibnamefont
  {{Cabrera-Lavers}}}, \ and\ \bibinfo {author} {\bibfnamefont
  {P.}~\bibnamefont {{Wroblewski}}},\ }\bibfield  {title} {\enquote {\bibinfo
  {title} {{The Distance, Mass, and Radius of the Neutron Star in 4U
  1608-52}},}\ }\href {\doibase 10.1088/0004-637X/712/2/964} {\bibfield
  {journal} {\bibinfo  {journal} {\apj}\ }\textbf {\bibinfo {volume} {712}},\
  \bibinfo {pages} {964--973} (\bibinfo {year} {2010})}\BibitemShut {NoStop}%
\bibitem [{\citenamefont {{{\"O}zel}}\ \emph {et~al.}(2012)\citenamefont
  {{{\"O}zel}}, \citenamefont {{Gould}},\ and\ \citenamefont
  {{G{\"u}ver}}}]{OzeGouGuv12}%
  \BibitemOpen
  \bibfield  {author} {\bibinfo {author} {\bibfnamefont {F.}~\bibnamefont
  {{{\"O}zel}}}, \bibinfo {author} {\bibfnamefont {A.}~\bibnamefont {{Gould}}},
  \ and\ \bibinfo {author} {\bibfnamefont {T.}~\bibnamefont {{G{\"u}ver}}},\
  }\bibfield  {title} {\enquote {\bibinfo {title} {{The Mass and Radius of the
  Neutron Star in the Bulge Low-mass X-Ray Binary KS 1731-260}},}\ }\href
  {\doibase 10.1088/0004-637X/748/1/5} {\bibfield  {journal} {\bibinfo
  {journal} {\apj}\ }\textbf {\bibinfo {volume} {748}},\ \bibinfo {eid} {5}
  (\bibinfo {year} {2012})}\BibitemShut {NoStop}%
\bibitem [{\citenamefont {{Suleimanov}}\ \emph {et~al.}(2011)\citenamefont
  {{Suleimanov}}, \citenamefont {{Poutanen}}, \citenamefont {{Revnivtsev}},\
  and\ \citenamefont {{Werner}}}]{SulPou11}%
  \BibitemOpen
  \bibfield  {author} {\bibinfo {author} {\bibfnamefont {V.}~\bibnamefont
  {{Suleimanov}}}, \bibinfo {author} {\bibfnamefont {J.}~\bibnamefont
  {{Poutanen}}}, \bibinfo {author} {\bibfnamefont {M.}~\bibnamefont
  {{Revnivtsev}}}, \ and\ \bibinfo {author} {\bibfnamefont {K.}~\bibnamefont
  {{Werner}}},\ }\bibfield  {title} {\enquote {\bibinfo {title} {{A Neutron
  Star Stiff Equation of State Derived from Cooling Phases of the X-Ray Burster
  4U 1724-307}},}\ }\href {\doibase 10.1088/0004-637X/742/2/122} {\bibfield
  {journal} {\bibinfo  {journal} {\apj}\ }\textbf {\bibinfo {volume} {742}},\
  \bibinfo {eid} {122} (\bibinfo {year} {2011})}\BibitemShut {NoStop}%
\bibitem [{\citenamefont {{Steiner}}\ \emph {et~al.}(2010)\citenamefont
  {{Steiner}}, \citenamefont {{Lattimer}},\ and\ \citenamefont
  {{Brown}}}]{SteLatBro10}%
  \BibitemOpen
  \bibfield  {author} {\bibinfo {author} {\bibfnamefont {A.~W.}\ \bibnamefont
  {{Steiner}}}, \bibinfo {author} {\bibfnamefont {J.~M.}\ \bibnamefont
  {{Lattimer}}}, \ and\ \bibinfo {author} {\bibfnamefont {E.~F.}\ \bibnamefont
  {{Brown}}},\ }\bibfield  {title} {\enquote {\bibinfo {title} {{The Equation
  of State from Observed Masses and Radii of Neutron Stars}},}\ }\href
  {\doibase 10.1088/0004-637X/722/1/33} {\bibfield  {journal} {\bibinfo
  {journal} {\apj}\ }\textbf {\bibinfo {volume} {722}},\ \bibinfo {pages}
  {33--54} (\bibinfo {year} {2010})}\BibitemShut {NoStop}%
\bibitem [{\citenamefont {{Abreu}}\ \emph {et~al.}(2007)\citenamefont
  {{Abreu}}, \citenamefont {{Hern{\'a}ndez}},\ and\ \citenamefont
  {{N{\'u}{\~n}ez}}}]{AbrHerNun07}%
  \BibitemOpen
  \bibfield  {author} {\bibinfo {author} {\bibfnamefont {H.}~\bibnamefont
  {{Abreu}}}, \bibinfo {author} {\bibfnamefont {H.}~\bibnamefont
  {{Hern{\'a}ndez}}}, \ and\ \bibinfo {author} {\bibfnamefont {L.~A.}\
  \bibnamefont {{N{\'u}{\~n}ez}}},\ }\bibfield  {title} {\enquote {\bibinfo
  {title} {{Sound speeds, cracking and the stability of self-gravitating
  anisotropic compact objects}},}\ }\href {\doibase
  10.1088/0264-9381/24/18/005} {\bibfield  {journal} {\bibinfo  {journal}
  {Classical and Quantum Gravity}\ }\textbf {\bibinfo {volume} {24}},\ \bibinfo
  {pages} {4631--4645} (\bibinfo {year} {2007})}\BibitemShut {NoStop}%
\end{thebibliography}
%

\end{document}